\documentclass[final,leqno,onefignum,onetabnum]{siamltex1213}

\usepackage{amssymb}
\usepackage[fleqn]{amsmath}
\usepackage{times}
\usepackage{enumerate}
\usepackage{subfigure}
\usepackage{multirow,extarrows}
\usepackage[arrow,matrix]{xy}
\usepackage{color}
\usepackage{stmaryrd}
\usepackage{bm}
\usepackage{ifpdf}
 \ifpdf
  \usepackage{epstopdf}
 \fi

\newcommand{\ds}{\displaystyle}
\newcommand{\dx}{\Delta x}
\newcommand{\dy}{\Delta y}
\newcommand{\dt}{\Delta t}

\newcommand{\rf}[1]{(\ref{#1})}

\newcommand{\vecx}{{\bf x}}

\renewcommand{\v}{{\bf v}}
\newcommand{\vecV}{{\bf V}}
\newcommand{\vecM}{{\bf M}}
\newcommand{\vecA}{{\bf A}}
\newcommand{\vecB}{{\bf B}}
\newcommand{\vecD}{{\bf D}}
\newcommand{\vecn}{{\bf n}}
\newcommand{\vecf}{{\bf f}}

\newcommand{\vech}{{\bf h}}
\newcommand{\vecH}{{\bf H}}
\newcommand{\vecK}{{\bf K}}

\newcommand{\vecg}{{\bf g}}

\newcommand{\vecP}{{\bf P}}
\newcommand{\vecO}{{\bf O}}
\newcommand{\vecu}{{\bf u}}

\newcommand{\vect}{{\bf t}}
\newcommand{\vsigma}{{\bf \sigma}}
\newcommand{\vz}{{\bf 0}}

\newcommand{\setR}{\mathbb{R}}

\newcommand{\p}{{+\frac{1}{2}}}

\newcommand{\jp}{{j+\frac{1}{2}}}
\newcommand{\jm}{{j-\frac{1}{2}}}
\newcommand{\ip}{{i+\frac{1}{2}}}
\newcommand{\im}{{i-\frac{1}{2}}}

\newcommand{\np}{{n+\frac{1}{2}}}

\renewcommand{\ij}{_{i,j}}
\newcommand{\ijp}{_{i,j}^{n+1}}
\newcommand{\ijn}{_{i,j}^{n}}

\title{Numerical schemes for the coupling of compressible and incompressible fluids in several space dimensions}

\author{Jochen Neusser\footnotemark[2]\ 
 \and Veronika Schleper\footnotemark[3]\ 
 }

\begin{document}
\maketitle

\renewcommand{\thefootnote}{\fnsymbol{footnote}}
\footnotetext[2]{\email{jochen.neusser@mathematik.uni-stuttgart.de}. Questions, comments, or corrections
to this document may be directed to that email address.}
\footnotetext[3]{\email{veronika.schleper@mathematik.uni-stuttgart.de}}
\renewcommand{\thefootnote}{\arabic{footnote}}

\begin{abstract}
We present a numerical scheme for immiscible two-phase flows with one compressible and one incompressible phase. Special emphasis lies in the discussion of the coupling strategy for compressible and incompressible Euler equations to simulate inviscid liquid-vapour flows. To reduce the computational effort further, we also introduce two approximate coupling strategies. The resulting schemes are compared numerically to a fully compressible scheme and show good agreement with these standard algorithm at lower numerical costs.
\end{abstract}



\pagestyle{myheadings}
\thispagestyle{plain}
\markboth{JOCHEN NEUSSER, VERONIKA SCHLEPER}{COUPLING OF COMPRESSIBLE AND INCOMPRESSIBLE FLUIDS}

\section{Introduction}
We consider a setting of two compressible, immiscible and isothermal fluids in a domain $\Omega\in\setR^d (d=1,2,3)$. The two fluids are separated by a sharp interface $\Gamma$ that does not contain mass. We refer to the fluid with the higher density as liquid and to that with the lower density as gas. If viscosity is neglected, the dynamics of the two fluids can be described by the compressible Euler equations

\begin{align}
&\begin{array}{l}
\partial_t\rho_g+\nabla\cdot(\rho_g \v_g)=0,\\[1.5ex]
\partial_t(\rho_g \v_g)+\nabla\cdot(\rho_g \v_g\otimes \v_g)+\nabla p_g(\rho_g)=0,
\end{array}&&\vecx\in\Omega_g(t),\label{gEuler}\\[2ex]
&\begin{array}{l}
\partial_t\rho_l+\nabla\cdot(\rho_l \v_l)=0,\\[1.5ex]
\partial_t(\rho_l \v_l)+\nabla\cdot(\rho_l \v_l\otimes \v_l)+\nabla p_l(\rho_l)=0,
\end{array}&&\vecx\in\Omega_l(t).\label{lEuler}
\intertext{The coupling between the liquid phase and the gas phase is modelled by the modified \textbf{Rankine-Hugoniot jump conditions}}
&\begin{array}{l}
(\v_g-\v_l)\cdot\vecn = 0,\\
p_l(\rho_l)-p_g(\rho_g)=(d-1)\sigma\kappa,
\end{array}&&\vecx\in\Gamma(t),\label{ccboundary}
\intertext{and a \textbf{kinetic relation}}
&\begin{array}{l}
\v_g\cdot\vecn = \gamma(t),
\end{array}&&\vecx\in\Gamma(t)\label{cckinetic_relation}.
\end{align}

Here, $\rho_g$ is the density of the gas, $\v_g$ is its velocity and  $p_g(\rho_g)$ is the pressure law, while $\rho_l,~\v_l$ and $p_l$ denote the analogous quantities for the liquid. 
$\gamma(t)$ is the normal velocity of the interface $\Gamma(t)$ and $\vecn(t)$ is the normal vector of $\Gamma(t)$ pointing into the liquid region. The kinetic relation \eqref{cckinetic_relation} and the coupling conditions \rf{ccboundary} ensure that the gas and the liquid have the same speed across the interface and they prevent any mass exchange between the two fluids. Note that we make use of the Young-Laplace law \cite{Buttbook} to take surface tension into account in \eqref{ccboundary}$_2$. Here $\kappa$ is the curvature of the interface and $\sigma>0$ is the (given) surface tension. 

The different behaviours of the gas phase and the liquid phase are taken into account by the equations of state $p_g,~p_l$. We assume the liquid to be only weakly compressible and we model its behavior by the Tait equation 
\begin{equation}
\label{tait}
p_{Tait}(\rho_l)=k_0\left(\left(\frac{\rho_l}{\rho_0}\right)^\gamma-1\right)+p_0,
\end{equation}
where $k_0$ depends on the Mach number and determines the compressibility of the fluid, $\rho_0$ and $p_0$ are density and pressure of the fluid at reference temperature at reference configuration and $\gamma>0$ is a constant exponent. On the other hand we describe the gas by the ideal gas law
\begin{equation}
\label{ideal}
p_{ideal}(\rho_g)=a^2\rho_g,
\end{equation}
where $a^2$ is the adiabatic gas constant.
The model \rf{gEuler}-\rf{ideal} is a standard model for the description of two compressible immiscible fluids \cite{FMW-11,HDW-13,Liu-Khoo-05,LeFloch-Thanh-02,Hu-09,MerkleRohde-07}, but it has a disadvantage from the numerical point of view. The equation of state for the liquid \eqref{tait} is very stiff and the speed of sound in the liquid region is very high. Numerically, this results in severe time step restrictions due to the CFL-condition. A different approach can be found in \cite{Quan-Schmidt-07} where two incompressible fluids are considered. However, this approach neglects some important properties of the two fluids. For the case that density changes in the gas phase cannot be neglected, Klein and Munz proposed a compressible/incompressible coupling, that is based on the introduction of multiple pressure variables \cite{Munzetal-15},\cite{Klein-95},\cite{Munzetal-03}.

Another way to circumvent the time step restrictions is to perform the compressible to incompressible limit only for the liquid and obtain a coupled system that is given by

\begin{align}
&\begin{array}{l}
\partial_t\rho_g+\nabla\cdot(\rho_g \v_g)=0,\\
\partial_t(\rho_g \v_g)+\nabla\cdot(\rho_g \v_g\otimes \v_g)+\nabla p_g(\rho_g)=0,
\end{array}&&\vecx\in\Omega_g(t),\label{cEuler}\\[2ex]
&\begin{array}{l}
\nabla\cdot\v_l=0,\\
\partial_t(\v_l)+\nabla\cdot( \v_l\otimes \v_l)+\frac{1}{\rho_l}\nabla p_l=0,
\end{array}&&\vecx\in\Omega_l(t),\label{iEuler}\\[2ex]
&\begin{array}{l}\v_g\cdot\vecn = \gamma(t)\end{array},&&\vecx\in\Gamma(t),\label{kinetic_relation}\\[2ex]
&\begin{array}{l}
(\v_g-\v_l)\cdot\vecn = 0,\\
p_l-p_g(\rho_g)=(d-1)\sigma\kappa,
\end{array}&&\vecx\in\Gamma(t).\label{boundary}
\end{align}

We recall that \eqref{gEuler}-\eqref{cckinetic_relation} and \eqref{cEuler}-\eqref{kinetic_relation} are known to be well posed at least in one space dimension, see \cite{Borsche-12},\cite{Colombo-12}. In the case of smooth solutions, there exist various well posedness results in higher space dimensions \cite{KlainermanMajda,Xu-11,Yong-05}

In this paper we develop a numerical algorithm for the coupled system \eqref{cEuler}-\eqref{boundary}. Up to our knowledge, such a scheme has not been proposed before. 
This contribution is structured as follows. 

Section \ref{ss:Riemann} explains the properties of the exact Riemann solver at the phase boundary, that is needed to construct physically correct solutions. 

Section \ref{s:Algorithm} contains the main outcome of this paper. We briefly introduce the numerical algorithms for the bulk phases, that are standard and can for example be found in \cite{Almgrenetal96,Torobook}. Then we give a detailed explanation of the coupling strategy we use at the phase boundary. This strategy is a modification of the fractional step scheme of  Bell, Colella and Glaz \cite{BCG-87}. Note that we rely on the ideas of the more recent paper \cite{Almgrenetal96}. We give analytical proof, that this coupling is well-defined. 

In Section \ref{s:numerics} we compare the resulting schemes  numerically to a fully compressible scheme. The results show good agreement with this standard algorithm at lower numerical costs.


\section{The Exact Riemann Solver}
\label{ss:Riemann}
\begin{figure}[h!]
\includegraphics[width=\textwidth]{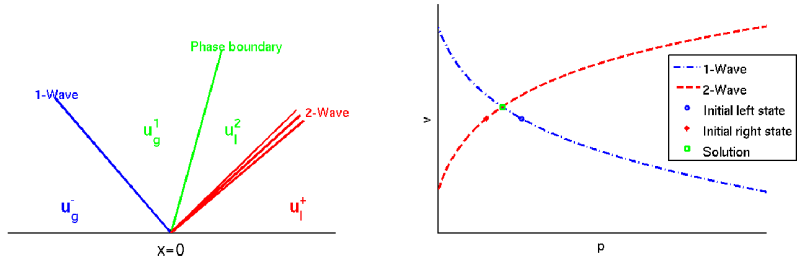}
  \caption{Illustration of the wave structure (left) and the geometric solution (right) of a Riemann problem at the interface.}
  \label{fig:lax_cc}
  \end{figure}

Consider again the fully compressible system \eqref{gEuler}-\eqref{cckinetic_relation}. In order to understand the coupling of the two fluids in the bulk phases across the interface we choose  $d=1$ for simplicity.  
At the interface, we need a thorough investigation of the wave structure in order to construct physically correct solutions. 
To this end, note that the speed of sound in the gas/liquid phase is defined as $c_i(\rho):=\sqrt{p_i(\rho_i)}$ for $i=g,l$. The eigenvalues and eigenvectors of the Jacobian of the one dimensional compressible Euler equations \rf{cEuler} are given by 
\begin{equation}
\begin{array}{ll}
\lambda_1=v_i-c_i(\rho_i),&\quad r_1=\left(\begin{array}{c}-1\\-v_i+c_i(\rho_i)
\end{array}\right),\\[2ex]
\lambda_2=v_i+c_i(\rho_i),&\quad r_2=\left(\begin{array}{c} 1\\ v_i+c_i(\rho_i)
\end{array}\right).
\end{array}
\end{equation}
for $i=g,l$.
Using the notation $\vecu_i:=(\rho_i, v_i)$, the corresponding one-wave curve through a given left state $u^-_i$ is defined by
\begin{equation}\label{Rleft}
\mathcal{L}^1_i(\sigma_1;u^-_i):=u^1_i(\sigma_1)
=\left\{\begin{array}{ll}
\left(\begin{array}{c}(1-\sigma_1)\rho^-_i\\v_i^- -\int_{\rho^-_i}^{(1-\sigma_1)\rho^-_i}\frac{c_i(r)}{r}dr
\end{array}\right),&\sigma_1\geq0\\
\left(\begin{array}{c}(1-\sigma_1)\rho^-_i\\v_i^--\sqrt{\left(\frac{1}{\rho^-_i}-\frac{1}{\rho^1_i(\sigma_1)}\right)\cdot\left(p(\rho^1_i(\sigma_1))-p(\rho^-_i)\right)}
\end{array}\right),&\sigma_1<0
\end{array}\right.
\end{equation}
Analogously, the (backward) two-wave curve through a given right state $u_i^+$ is defined by
\begin{equation}\label{Rright}
\mathcal{L}^2_i(\sigma_2;u^+_i):=u^2_i(\sigma_2)
=\left\{\begin{array}{ll}
\left(\begin{array}{c}(1+\sigma_2)\rho^+_i\\v_i^++\int_{\rho_i^+}^{(1+\sigma_2)\rho^+_i}\frac{c_i(r)}{r}dr
\end{array}\right),&\sigma_2\geq0\\
\left(\begin{array}{c}(1+\sigma_2)\rho^+_i\\v^+_i-\sqrt{\left(\frac{1}{\rho^+_i}-\frac{1}{\rho^2_i(\sigma_2)}\right)\cdot\left(p(\rho^2_i(\sigma_2))-p(\rho^+_i)\right)}
\end{array}\right),&\sigma_2<0
\end{array}\right.
\end{equation}
The detailed derivation of these wave curves can be found in \cite{Dafermosbook}. Figure \ref{fig:lax_cc} illustrates the wave structure and the geometric solution to the \textbf{Riemann problem} at the interface.
Without loss of generality, let the interface be given at $x=0$ with the gas phase $u^-_g$ in $\mathbb{R}^-$ and the liquid phase $u_l^+$ in $\setR^+$. We consider the Riemann problem in the gas phase with initial conditions
\begin{equation}
u^0=\left\{\begin{array}{ll}
u^-_g,&\quad x\in\setR^-,\\
u^+_l,&\quad x\in\setR^+.
\end{array}\right.
\end{equation}
In order to solve the Riemann problem, we have to find $\sigma_1,\sigma_2$ defined in equations \rf{Rleft},\rf{Rright} such that
\begin{equation}
\begin{array}{l}
v^1_g(\sigma_1)=v^2_l(\sigma_2),\\
p_g(\rho_g^1(\sigma_1))=p_l(\rho_l^2(\sigma_2)).
\end{array}
\end{equation}
The unique solvability of this Riemann problem at least locally is well-known in the literature and can be checked by the inverse function theorem. We refer to \cite{Colombo-12} for detailed analytical results. 

We now consider the Riemann problem at the interface $\Gamma$ for the compressible-incompressible system \eqref{cEuler}-\eqref{boundary}. 

In contrast to the coupling between two compressible fluids, we cannot use the approach we outlined above, since the Riemann problem at the interface will not lead to the right solution because of the special structure of Riemann problems. The fact that the liquid part in the Riemann problem fills the whole halfspace $\setR^+$ implies that the incompressible phase has infinite mass. Therefore, any incoming wave of the compressible phase will be completely reflected by the incompressible fluid phase. Thus, the local Riemann problem cannot give the physically right solution in the context of coupled compressible-incompressible flows with finite liquid mass. Instead, we have to consider the whole liquid phase explicitly in order to impose the finite mass of the incompressible phase. 

In one space dimension, the velocity of the liquid is constant and can be obtained by the ordinary differential equation
\begin{equation}
v_l(t)=\ds\frac{p_g^-(t)-p_g^+(t)}{L \rho_l},
\end{equation}
where $L$ is the length of the droplet and $p_g^-~(p_g^+)$ are the values left (right) of the liquid.
In two space dimensions, on has to solve the fully coupled problem, that is given by the incompressible constraint
$$\nabla\cdot\v_l=0,$$
and the coupling conditions
\begin{equation}
\begin{array}{l}
\v_g(\sigma)=\v_l,\\
p_g(\rho_g(\sigma))=p_l,
\end{array}
\end{equation}
where $\rho_g(\sigma),\v_g(\sigma)$ are obtained from the Lax curves \eqref{Rleft},\eqref{Rright}.
Note that some difficulties for this coupling arise in two and three space dimensions. 
If we for example consider a liquid droplet that is surrounded by gas, the interface condition has to be fulfilled for all $\vecx\in\Gamma$ simultaneously (cf. Figure \ref{fig:1setting}).
We have to determine the boundary states to the left and right of the interface $\Gamma(t)$ such that the incompressibility condition $\nabla\cdot\v_l=0$ is fulfilled in $\Omega_l$ and such that the new boundary states in the gas phase are admissible for the compressible model. 
This task is not trivial and we refer to Section \ref{s:coupling} for a detailed solution strategy on the discrete level.

\section{The Numerical Algorithm for the Coupled System}
\label{s:Algorithm}
In this section we present a numerical scheme to solve the coupled compressible/incompressible system \eqref{cEuler}-\eqref{boundary}. 

We make some simplifications concerning the geometry and discretization of the domain $\Omega$ because we want to focus on the treatment of the incompressible/compressible coupling.
\begin{figure}
\centering
\includegraphics[width=.5\columnwidth]{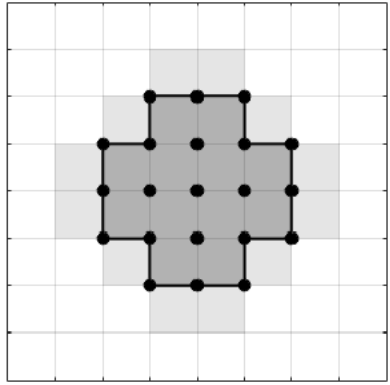}
\caption{Discretization of the domain $\Omega$ in two space dimensions. A liquid droplet (light grey) is separated from the gas phase by a sharp interface (bold line).}
\label{fig:1setting}
\end{figure}
We assume that we have a rectangular domain $\Omega$ that is discretized by Cartesian mesh with rectangular cells $B_{i,j}$ (cf. Figure \ref{fig:1setting} for an example of such a domain). Furthermore, we assume that each cell $B_{i,j}$ is either contained in the compressible region $\Omega_g$ (Fig. \ref{fig:1setting} white/light grey) or in the incompressible region $\Omega_l$ (Fig. \ref{fig:1setting} dark grey). This means that the phase boundary (cf. Fig. \ref{fig:1setting} bold line) is given by a finite set of edges. We define $\mathbb{B}=\{B_{i,j}|B_{i,j}\in\Omega\}$ as the discrete version of $\Omega$ and
\[\mathbb{B}_l=\{B_{i,j}|(i,j)\in\mathcal{I}_l\},\quad\mathbb{B}_g=\{B_{i,j}|(i,j)\in\mathcal{I}_g\},\]
as the discrete versions of $\Omega_l,~\Omega_g$ respectively. Here $\mathcal{I}_l$ is the list of all cell indices in the in liquid phase and $\mathcal{I}_g$ is the list of all cell indices in the in gas phase.
Additionally, we denote the list of all nodes in the liquid region by
$\mathcal{I}^{nodes}_l$ (Fig. \ref{fig:1setting} black dots). The light grey rectangles in Figure \ref{fig:1setting} mark the elements of the compressible phase that are needed for the interface treatment.\\
In the following sections, we specify in detail the bulk phase algorithms for the compressible and incompressible phases.
%
\subsection{Discretization of the compressible Euler equations}\label{ss:compressible}
In this section, we give a brief overview of the numerical scheme for the compressible Euler equations. The steps of the scheme are standard but are presented for the sake of completeness. Note that we set $\vecu=\vecu_g$ in this section to avoid overloaded notations.
We are concerned with solving numerically the general initial boundary value problem
\begin{equation}
\begin{array}{lcl}
\vecu_t+\vecf(\vecu)_x+\vecg(\vecu)_y=0&,&\vecx\in\Omega_g\\
\vecu(\vecx,0)=\vecu^0(\vecx)&,&\vecx\in\Omega_g\\
\vecu(\vecx,t)=\vecu_b(t)&,&\vecx\in\partial\Omega_g
\end{array}
\end{equation}
with $\vecx=(x,y)^T,$ $\vecu=(\rho,\rho u,\rho v)^T$, $\vecf=(\rho u,\rho u^2+p,\rho uv)^T$ and $\vecg=(\rho v,\rho uv,\rho v^2+p)$. We use the explicit conservative formula
\begin{equation}
\vecu\ijp=\vecu\ijn-\frac{\dt}{\dx}[\vecf_{\ip,j}-\vecf_{\im,j}]-\frac{\dt}{\dy}[\vecg_{i,\jp}-\vecg_{i,\jm}],\quad (i,j)\in\mathcal{I}_g
\end{equation}
where the numerical fluxes $\vecf_{\ip,j},~\vecg_{i,\jp}$ are the HLL fluxes by Harten, Lax and van Leer \cite{HLL-83}, see also \cite{Torobook}.
Using the Roe averaged values \cite{Roe-81}
\begin{equation}
\tilde{\rho}_{\ip,j}=\ds\sqrt{\rho_{i,j}\rho_{i+1,j}},\quad\tilde{u}_{\ip,j}=\frac{\sqrt{\rho_{i,j}}u_{i,j}+\sqrt{\rho_{i+1,j}}u_{i+1,j}}{\sqrt{\rho_{i,j}}+\sqrt{\rho_{i+1,j}}}
\end{equation}
to compute the speeds on both sides of the interface
\begin{equation}
S^-=\tilde{u}_{\ip,j}-\sqrt{p'(\tilde{\rho}_{\ip,j})},\quad S^+=\tilde{u}_{\ip,j}+\sqrt{p'(\tilde{\rho}_{\ip,j})},
\end{equation}
we obtain the HLL numerical flux 
\begin{equation}
\vecf_{\ip,j}=\left\{\begin{array}{cccc}
\vecf(\vecu_{i,j}),&\text{if}&0\leq S^-,\\[2ex]
\ds\frac{S^+\vecf(\vecu_{i,j})-S^-\vecf(\vecu^+)+S^-S^+(\vecu_{i+1,j}-\vecu_{i,j})}{S^+-S^-},&\text{if}&S^-<0<S^+,\\[2ex]
\vecf(\vecu_{i+1,j}),&\text{if}&0\geq S^+,
\end{array}\right.
\end{equation}
An analogous result can be obtained for $\vecg_{i,\jp}$ by considering the Riemann problem in $y$-direction.

This method is an explicit finite volume scheme and it requires a time step restriction. 
The standard stability analysis shows that we must require
\begin{equation}
\label{cflcomp}
\dt=\sigma_{c}\ds\min_{i,j}\left(\frac{\dx}{(|\tilde{u}_{\ip,j}|+p(\tilde{\rho}_{\ip,j}))},\frac{\dy}{(|\tilde{v}_{i,\jp}|+p(\tilde{\rho}_{i,\jp}))}\right),
\end{equation}
where $\sigma_{c}<1$ is the CFL number.

\begin{figure}[!ht]
\centering
\includegraphics[width=.5\linewidth]{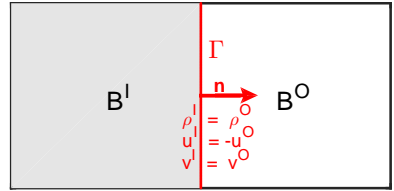}
\caption{A compressible cell $B_I$ (grey) is separated from a boundary cell $B_O$ (white) by the interface $\Gamma$ (bold red line). The boundary conditions are given by a reflecting wall (a)).}
\label{fig:boundary}
\end{figure}

We complete the discretization with the description of the boundary conditions. Here $B_I$ denotes a cell in the compressible domain that is separated from a boundary cell $B^O$ by the interface $\Gamma$. Figure \ref{fig:boundary} displays the situation for a vertical boundary.
We prescribe values $\vecu^O$ in these boundary cells to enforce the boundary conditions. Denoting by n the outward normal vector and by t the tangential vector, the  different boundary conditions and the corresponding boundary cell values are given by\\
\begin{itemize}
\item[a)] Reflecting wall: $\v(\vecx)\cdot\vecn=0,~\vecx\in\partial\Omega_g$
{\setlength{\mathindent}{2cm}
\[\rho^O=\rho^I,\quad \v^O\cdot\vecn=-\v^I\cdot\vecn,\quad \v^O\cdot\vect=\v^I\cdot\vect.\]
\item[b)] Inflow: $\vecu=\vecu^{inflow}(\vecx),~\vecx\in\partial\Omega_g$
\[\vecu=\vecu^{inflow}.\]
\item[c)] Outflow:
\[\vecu^O=\vecu^I.\]
\item[d)] Compressible-incompressible interface: 
\[p^O=p^{Inc},\quad \v^O\cdot\vecn=\v^{Inc}\cdot\vecn,\quad \v^O\cdot\vect=\v^I\cdot\vect.\]}
$\rho^{Inc},~\v^{Inc}$ are obtained from the algorithm for the incompressible equations, that is presented in Section \ref{s:coupling}. This is a modification of the ghost-fluid method by Fedkiw et al. \cite{Fedkiw-99}.
\end{itemize}

\subsection{Discretization of the incompressible Euler equations}
In this section, we propose an algorithm for the incompressible Euler equations \rf{iEuler}, that incorporates the compressible-incompressible coupling conditions \rf{boundary}.  

Ideally we would like to compute
\begin{equation}\label{ieulersemi}
\begin{array}{l}
\ds\frac{\v^{n+1}-\v^n}{\dt}+[\nabla\cdot( \v\otimes \v)]^{n+1}+\frac{1}{\rho_l}\nabla p^{n+1}=0,\\[1.5ex]
\nabla\cdot\v^{n+1}=0.
\end{array}
\end{equation}
Therefore, we use a modification of the fractional step scheme, that was introduced by Bell, Colella and Glaz \cite{BCG-87}. There are various other numerical methods available for the incompressible Euler equations, for example the projection method that was first brought up by Chorin \cite{Chorin-68a,Chorin-69}. Primitive-variable methods that rely on a staggered grids can be found in \cite{HarlowWelch-65,KimMoin-80}. The streamline diffusion method was introduced in \cite{BrooksHughes-82} and uses space-time finite elements.\\
Our scheme relies particularly on the ideas of the more recent paper \cite{Almgrenetal96} and consists of two steps that we recall here for convenience. Note that the second step will be modified below to account for the coupling with the compressible phase. In the first step we solve the convection equation \eqref{fractional1} at time level $t^\np$ without enforcing the incompressibility constraint. The second step is a projection that takes the coupling between the compressible and the incompressible fluid into account and imposes the incompressibility constraint.
For the convection step we solve
\begin{equation}\label{fractional1}
\frac{\v^\np-\v^n}{\dt}+[\nabla\cdot( \v\otimes\v)]^n=0
\end{equation}
for the intermediate velocity $$\v^\ast:=\ds\frac{\v^\np-\v^n}{\dt}.$$
The convection terms  at $t^\np$ are explicitly computed from the velocity data at $t^n$ using a first order upwind scheme. The resulting velocity field $\v^\ast$ is in general not divergence free. 

The projection step decomposes the result of the first step into an approximately divergence-free vector field and a discrete gradient of a scalar potential. They correspond to the update for the velocity and to the new approximation for the pressure, respectively.
Denoting by $\mathcal{P}$ the projection operator, we obtain $\v^{n+1},p^{n+1}$ by 
\begin{equation}\label{fractional2}
\begin{array}{l}
\ds\frac{\v^{n+1}-\v^n}{\dt}=\mathcal{P}\left(\v^\ast\right),\\[2ex]
\ds\frac{1}{\rho_l}\nabla p^{n+1}=(\mathcal{I}-\mathcal{P})\left(\v^\ast\right).
\end{array}
\end{equation} 


\subsubsection{Discretization of the convection step}
In this section we recall the algorithm for the convection step that was proposed in \cite{Almgrenetal96}. In contrast to them, we use a first order finite volume upwind scheme for the inviscid Burgers equation \eqref{fractional1}. This is sufficient in the present context since the coupling of the compressible and incompressible fluid will also be only first order accurate. Here $\v\ijn$ represents the value of the velocity field in cell $B_{ij}$ at time $t^n$. Note that we present the results in this section for convenience and one can find a more detailed explanation in \cite{Almgrenetal96}.
We have 
\begin{equation}
\ds\frac{\v^\np-\v^n}{\dt}+
\vech(\v^n)_x+\vecK(\v^n)_y=0
\end{equation}
with $\vech(\v^n)=(u^n u^n,u^n v^n),~\vecK(\v^n)=(u^n v^n,v^n v^n)$.

In order to obtain a discrete scheme, we first define the normal advective velocity
\begin{equation}
u_{\ip,j}^{adv}=\left\{\begin{array}{ll}u^n_{i,j}\quad &\text{if}~u^n_{i,j}>0,\quad u^n_{i,j}+u^n_{i+1,j}>0,\\0\quad &\text{if}~u^n_{i,j}\leq0,\quad u^n_{i+1,j}\geq 0 \\u^n_{i+1,j}\quad &\text{otherwise}\end{array}\right.
\end{equation}
We now upwind $\v^n_{\ip,j}$ based on $u^{adv}_{\ip,j}$:
\begin{equation}
\v^n_{\ip,j}=\left\{\begin{array}{cl}\v^n_{i,j}\quad &\text{if}~u^{adv}_{\ip,j}>0,\\\frac{1}{2}(\v^n_{i,j}+\v^n_{i+1,j})\quad &\text{if}~u^{adv}_{\ip,j}=0,\\\v^n_{i+1,j}\quad &\text{otherwise}\end{array}\right.
\end{equation}
After constructing $\v_{\im,j},~\v_{i,\jp},~\v_{i,\jm}$ in a similar manner, we use these upwind values to compute the numerical fluxes:
\begin{equation}
\begin{array}{l}
\vecH_{\ip,j}^n=(u_{\ip,j}^n u_{\ip,j}^n,u_{\ip,j}^n v_{\ip,j}^n)^T,\\
\vecK_{i,\jp}^n=(u_{i,\jp}^n v_{i,\jp}^n,v_{i,\jp}^n v_{i,\jp}^n)^T.
\end{array}
\end{equation}
It remains to define the boundary values, that come from the compressible/incompressible coupling. Without any loss of generality, we assume that $B_{i,j}$ is an incompressible cell and $B_{i+1,j}$ is a compressible cell. We set
\begin{equation}
\v^n_{i+1,j}=\left(\begin{array}{c}
u^n_{i+1,j}\\v^n_{i,j}
\end{array}\right).
\end{equation}
Here $u^n_{i+1,j}$ is the velocity in the compressible cell that we computed in the previous time step.

With these fluxes we can define the upwind scheme
\begin{equation}
\ds\frac{\v_{i,j}^\np-\v\ijn}{\dt}+\frac{\vecH_{\ip,j}^n-\vecH_{\im,j}^n}{\dx}+\frac{\vecK_{i,\jp}^n-\vecK_{i,\jm}^n}{\dy}=0.\quad (i,j)\in\mathcal{I}_l
\end{equation}

The Godunov method is an explicit finite volume scheme, and requires the time step restriction.

\begin{equation}
\dt=\sigma_{inc}\ds\min_{ij}\left(\frac{\dx}{|u_{ij}|},\frac{\dy}{|v_{ij}|}\right),
\end{equation}
where $\sigma_{inc}<1$ is the CFL number. Note that this time step restriction is severely weaker than the time step restriction in the compressible phase \eqref{cflcomp}, since it does not involve the speed of sound.

\subsubsection{Discretization of the projection step}
\label{Projection}
In the previous step we computed the intermediate velocity $\v^\ast$. We already mentioned that this vector field is not in general divergence free.  In the projection step, we again rely on the ideas of \cite{Almgrenetal96}. We recall the basic concepts of this step here to fix the ideas. In the next section, this step will be modified to include the coupling of compressible and incompressible phase. We decompose $\v^\ast$ into a discrete divergence-free vector field and a gradient field. This decomposition is based on a finite element formulation for the pressure $p$.

We denote by $\mathcal{V}^{r,s}_h$ the space of element wise defined polynomials of degree $r$ in $x$-direction and $s$ in $y$-direction:
\begin{equation}
\mathcal{V}^{r,s}_h=\ds\left\{\phi_h~\Big|~\phi_h|_{_{B_{i,j}}}\in\mathbb{P}^{r,s}(B_{i,j}),~ \forall(i,j)\in\mathcal{I}_l\right\},
\end{equation}
where $\mathbb{P}^{r,s}(B_{i,j})$ is the space of polynomials of degree $r$ in $x$-direction and $s$ in $y$-direction on the cell $B_{i,j}$.

We consider the scalar pressure field $p$ to be a continuous bilinear function over each cell, i.e., the pressure is in 
\begin{equation}
S_h=\mathcal{V}^{1,1}_h\cap C^0(\Omega_l).
\end{equation}
We define $\Psi$ as the standard basis for $S^h$, namely
\begin{equation}
\Psi=\left\{\ds\psi_{i+\frac{1}{2},{j+\frac{1}{2}}}(\vecx)\left|\psi_{{i+\frac{1}{2}},{j+\frac{1}{2}}}(\vecx_{{i'+\frac{1}{2}},{j'+\frac{1}{2}}})=\delta_{ii'}\delta_{jj'},\forall (i,j),~(i',j')\in\mathcal{I}^{nodes}_l\right.\right\}.
\end{equation}
Here $\vecx_{i\pm\frac{1}{2},j\pm\frac{1}{2}}$ are defined at the nodes of the cell $B_{i,j}$.

For the velocity space we define
\begin{equation}
\vecV_h=\mathcal{V}^{0,1}_h\times\mathcal{V}^{1,0}_h,
\end{equation}
thus $u$ is piecewise constant in $x$ and a discontinuous linear function in $y$ in each cell and $v$ is piecewise constant in $y$ and a discontinuous linear function in $x$ in each cell.


Note that $\v^\ast$, computed in the previous step, is given by cell averages, while the velocity space $\vecV_h$ is larger and allows not only piecewise constant functions, but also functions with linear variations. This allows us to define the projection step. See \cite{Almgrenetal96} for more details.

Thus we are able define a decomposition of $\vecV_h$ into two orthogonal components
\begin{equation}\label{orthogonaldecomposition}
\vecV_h=\overline{\vecV}_h\oplus\vecV_h^{\bot}
\end{equation}
where $\overline{\vecV}_h$ represents the cell average and $\vecV_h^\bot$ represents the orthogonal linear variation. In particular, for each $\v\in\vecV_h$ we define the cell average by
$$\overline{\v}_{i,j}=\frac{1}{|B_{i,j}|}\int_{B_{i,j}}\v d\vecx$$ and the orthogonal linear variation by 
$$\v_{i,j}^{\bot}=\v_{i,j}-\overline{\v}_{i,j}.$$ 
A similar decomposition of $\nabla p$ can be defined for all $ p\in S^h$ by
\begin{equation}
(\nabla p)\ij=(\overline{\nabla p})\ij+(\nabla p)\ij^\bot.
\end{equation}
With these preparations, we can define the projection step in \rf{fractional2}. 

We only demand the new velocity field $\v^{n+1}$ to be weakly divergence free in \eqref{ieulersemi}. Therefore we define a vector field $\v^d\in\vecV_h$ to be weakly divergence free if
\begin{equation}
\ds\int_\Omega \v^d\cdot\nabla\psi d\vecx=0,\quad\forall\psi\in S_h.
\end{equation}
We use this definitions to introduce a decomposition of the vector field $\v^\ast\in\vecV^h$ onto a gradient $\nabla p^{n+1}$ and weakly divergence-free field $\v^d$.  
We rewrite \eqref{ieulersemi}$_1$ by using \rf{fractional1} and then apply the divergence operator to obtain
\begin{equation}\label{div}
\nabla\v^d=\nabla\v^\ast-\frac{1}{\rho_l}\Delta p^{n+1}.
\end{equation}
Then we write \rf{div} in its weak form, do partial integration and obtain \begin{equation}\label{weak1}
\begin{array}{l}
\ds\int_\Omega(\v^\ast-\v^d)\cdot\nabla\psi d\vecx
+\ds\int_{\partial\Omega}(\v^d-\v^\ast)\cdot(\psi\vecn)ds\\[2.5ex]
=\ds\frac{1}{\rho_l}\int_\Omega\nabla p^{n+1}\cdot\nabla\psi d\vecx
-\frac{1}{\rho_l}\int_{\partial\Omega}\nabla p^{n+1}\cdot(\psi\vecn)ds
\end{array}\forall\psi\in S_h.
\end{equation} 
As we want $\v^d$ to be weakly divergence-free, we have to solve 
\begin{equation}\label{weak2}
\begin{array}{l}
\ds\int_\Omega\v^\ast\cdot\nabla\psi_{i+\frac{1}{2},j+\frac{1}{2}} d\vecx+\ds\int_{\partial\Omega}(\v^d-\v^\ast)\cdot(\psi_{i+\frac{1}{2},j+\frac{1}{2}}\vecn)ds\\[2.5ex]
=\ds\frac{1}{\rho_l}\int_\Omega\nabla p^{n+1}\cdot\nabla\psi_{i+\frac{1}{2},j+\frac{1}{2}} d\vecx-\frac{1}{\rho_l}\int_{\partial\Omega}\nabla p^{n+1}\cdot(\psi_{i+\frac{1}{2},j+\frac{1}{2}}\vecn)ds
\end{array}\forall\psi_{i+\frac{1}{2},j+\frac{1}{2}}
\end{equation} 
for 
$$p^{n+1}(\vecx)=\ds\sum_{i,j}p^{n+1}_{i+\frac{1}{2},j+\frac{1}{2}}\psi_{i+\frac{1}{2},j+\frac{1}{2}}(\vecx).$$
Here the $\psi_{i\p,j\p}$ are the standard basis functions for $S_h$ as defined above. 

Note that the projection is applied to the vector field $\overline{\v}^\ast$ instead of $\v^\ast$, so we make the implicit assumption that $\v^{\bot,\ast}=0$.
With the updated $p^{n+1}$ we compute 
\begin{equation}
\ds\overline{\v}^d=\overline{\v}^\ast-\ds\frac{1}{\rho_l}\overline{\nabla p} ^{n+1}
\end{equation}
as the approximation of $(\v^{n+1}-\v^n)/\dt$ in \rf{fractional2}.
The vector field $\overline{\v}^d$ is only approximately divergence free; i.e., 
$\ds\int_\Omega \overline{\v}^d\cdot\nabla\psi d\vecx\neq0$. The quantity that is weakly divergence free is 
$$\v^d=\v^\ast-\ds\frac{1}{\rho_l}p^{n+1}.$$ 
However, it can be shown that we have $D\overline{\v}^{n+1}=\mathcal{O}(\dx)$, where the discrete divergence $D\v$ is defined as 
$$(D\v)_{i\p,j\p}=-\ds\frac{1}{|\psi|_\Omega}\int_\Omega\v\cdot\nabla\psi_{i\p,j\p}dx
$$
with $\ds|\psi|_\Omega=\int_\Omega\psi_{i\p,j\p}dx$, see \cite{Almgrenetal96}. 

To accomplish this, we rewrite the result of the decomposition \rf{fractional2} as
\begin{equation}\label{decomposition}
\begin{array}{ll}
\ds\frac{\v^\ast-\v^n}{\dt}=\frac{\overline{\v}^\ast-\overline{\v}^n}{\dt}-\frac{1}{\rho_l}p^{n+1}\\
D\v^{n+1}=0.
\end{array}
\end{equation}
where we implicitly assume $D\v^n=0$. Since $D\overline{\v}^{n+1}=-D\v^{\bot,n}$, we can estimate how errors behave by examining the behaviour of $\v^{\bot,n}$. Therefore we break \rf{decomposition} into its component pieces in :
\begin{equation}\label{decomposition2}
\begin{array}{ll}
\ds\frac{\overline{\v}^{n+1}-\overline{\v}^n}{\dt}+\frac{\v^{\bot,n+1}-\v^n}{\dt}=\frac{\overline{\v}^\ast-\overline{\v}^n}{dt}-\frac{1}{\rho_l}\nabla p^{n+1}.\\
\end{array}
\end{equation}
By using the definition of $\vecV_h$ given by \rf{orthogonaldecomposition}, we obtain
\[\overline{\v}^{n+1}=\overline{\v}^n+\dt\left(\frac{\overline{\v}^\ast-\overline{\v}^n}{\dt}-\frac{1}{\rho_l}\overline{\nabla p} ^{n+1}\right)\]
and 
\[\v^{\bot,n+1}=\v^{\bot,n}-\dt(\nabla p^{n+1})^\bot\approx\v^{\bot,n}-\dt(\nabla p^{n})^\bot-\dt^2(\nabla p^{n}_t)^\bot.\]
As long as the pressure $p$ remains sufficiently smooth in space and time, $\v^\bot$ is well behaved and scales with $\dt=\sigma_{inc} \dx$. Thus $D\v^n=0$ guarantees that $\overline{\v}^{n+1}=\mathcal{O}(\dx)$ as long as the CFL condition is enforced. 

\subsubsection{Implementation of the coupling projection}
\label{s:coupling}
It remains to include the coupling between the two fluids into the projection scheme \eqref{weak2}. This coupling is the key ingredient to our numerical scheme and it relies on the choice of the approximate Riemann solver. In Section \ref{ss:Riemann} we introduced the exact Riemann solver and we use this solver to outline the solution strategy at the phase boundary. We want to point out that this solver can be replaced by an approximate one. We refer to the end of this section for further explanations and examples for approximate solvers in this context.
In order to solve the coupling correctly, the boundary states $\rho_g^{n+1},~\v_g^{n+1}\cdot\vecn,\v_l^{n+1}\cdot\vecn~\text{and}~p_l^{n+1}$ have to fulfil the interface conditions \eqref{boundary} simultaneously. Therefore we have to determine these boundary states such that $(\rho_g^{n+1},\v_g^{n+1}\cdot\vecn)=\mathcal{L}^1_g(\sigma;(\rho_g^n,\v_g^n\cdot\vecn))$ for some $\sigma$ and $\v_g^{n+1}\cdot\vecn=\v_l^{n+1}\cdot\vecn,~p_l^{n+1}=p_g(\rho_g^{n+1})$.
Before we show formally that this is possible, we outline the discretization for the projection

\begin{equation}\label{discreteprojectionboundary}
\begin{array}{ll}
&\ds\int_\Omega \v^\ast\cdot\nabla\psi_{i\p,j\p} d\vecx+\int_{\partial\Omega}(\v^d-\v^\ast)\cdot(\psi_{i\p,j\p}\vecn)ds\\[2.5ex]
=&\ds\frac{1}{\rho_l}\int_\Omega\nabla p^{n+1}\cdot\nabla\psi_{i\p,j\p} d\vecx-\frac{1}{\rho_l}\int_{\partial\Omega}\nabla p^{n+1}\cdot(\psi_{i\p,j\p}\vecn)ds
\end{array}
\end{equation}

\begin{figure}[ht!]
\centering
\includegraphics[width=.75\columnwidth]{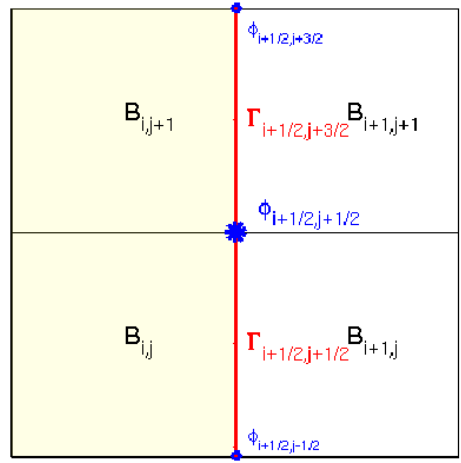}
\label{fig:interface}
\caption{Coupling between the incompressible (yellow) and compressible domain (white) across the interface (bold red). }
\end{figure}
We now take a closer look at the interface between two liquid cells $B^l_{i,j},B^l_{i,j+1}$ and two gas cells $B^g_{i+1,j},B^g_{i+1,j+1}$ to describe how the interface coupling works in detail. Therefore we consider the situation that is illustrated in Figure \ref{fig:interface}.  The coupling for $\v^{n+1}$ is straightforward as $\v^{n+1}$ is constant along cell boundaries. We obtain after a brief calculation
\begin{equation}\label{bc_velocity}
\begin{array}{ll}
\ds\int_\Gamma (\v^d-\v^\ast)\cdot(\psi_{i\p,j\p}\vecn)~ ds=&\ds\frac{\dy}{2\dt}(u_{i,j+1}^{n+1}(\vsigma_{i\p,j\p})-u^\ast_{i,j+1}\\[2ex]
&\ds+u_{i+1,j+1}^{n+1}(\vsigma_{i\p,j+\frac{3}{2}})-u^\ast_{i+1,j+1}).
\end{array}
\end{equation}
The coupling for $p^{n+1}$ is not so straightforward as $p_{i\p,j\p}^{n+1}$ is defined at cell nodes and is not constant along cell boundaries. We incorporate the interface condition by setting 
\begin{equation}\label{bc_pressure}
p^{n+1}_{i\p,j\p}=p(\rho_{i,j+1}^{n+1}(\vsigma_{i\p,j\p})).
\end{equation}
in the third term in \rf{discreteprojectionboundary}. This is the standard treatment for Dirichlet boundary conditions in elliptic equations.

In other words, we substitute the interface quantities $\v^{n+1},p^{n+1}$ by the one parameter families $\mathcal{L}_g^1(\vsigma;\rho_g^n,\v_g^n\cdot\vecn)$. Note that this substitution does not chance the number of unknowns in \rf{discreteprojectionboundary}.

At inner cell nodes we set $p^{n+1}_{i\p,j\p}=\vsigma_{i\p,j\p}$ to have a consistent notation in the next steps.

We put the discretizations for $\v^d,\v^\ast,p^{n+1}$ together and obtain the solution operator $\vecO(\vsigma)$, that is defined as
\begin{equation}\label{operator}
\begin{array}{ll}
\vecO(\vsigma)=&\ds\frac{1}{\rho_l}\int_{\Omega_l}\nabla p^{n+1}(\vsigma)\cdot\nabla\psi_{i\p,j\p} d\vecx\\
[2ex]
&\ds -\int_{\Omega_l}\v^\ast\cdot\nabla\psi_{i\p,j\p} d\vecx\\[2ex]
&\ds-\int_\Gamma (\v^d-\v^*)(\vsigma)\cdot(\psi_{i\p,j\p}\vecn)~ ds,
\end{array}
\quad\forall (i+1/2,j+1/2)\in\mathcal{I}_l^{nodes}
\end{equation}
We have to check if \rf{operator} is locally solvable, in order to proof that the interface coupling for our scheme is well defined. 

\begin{theorem}
\label{maintheorem}
Let $\vecO:\setR^M\to\setR^M$ be defined as in \eqref{operator}. Then $\vecO(\vsigma)$ satisfies
\begin{equation}\label{det}
\det\left[~D_{\vsigma_1}\vecO({\bf 0})~D_{\vsigma_2}\vecO({\bf 0})~\dots~D_{\vsigma_N}
\vecO({\bf 0})~\right]\neq0.
\end{equation}
\end{theorem}

Proof: Without any loss of generality, we assume that the nodes $K:=(i+1/2,j+1/2)\in\mathcal{I}^{nodes}_l$ are sorted in the following way. The first $M$ nodes $K_1,\dots,K_M$ are the boundary nodes in clockwise order and the remaining $N-M$ nodes $K_{M+1},\dots,K_N$ are the inner nodes. We want to point out, that for a free droplet without wall contact the number of boundary edges corresponds to the number of boundary nodes. 
We take a closer look at the structure of the solution operator \rf{operator} and see that the second term does not depend on $\vsigma$ and so we do not have to take it into consideration for the computation of the determinant. The first term can be rewritten as
\[
\frac{1}{\rho_l}\int_{\Omega_l}\nabla p^{n+1}(\vsigma)\cdot\nabla\psi_K d\vecx=\frac{1}{\rho_l}{\bf MP}^{n+1}(\vsigma)
\]
with $\bf{M}$ the mass matrix defined as
\begin{equation}
m_{k,l}:=\left\{\begin{array}{ll}
\ds\delta_{k,l}&\vecx_{K_k}\in\Gamma\\
\ds\int_{\Omega_l}\nabla\psi_{K_k}\cdot\nabla\psi_{K_l}d\vecx&\text{otherwise}
\end{array}\right.
\end{equation}
and ${\bf P}^{n+1}(\vsigma)$ defined as
\begin{equation}\label{Pressure}
\begin{array}{l}
{\bf P}^{n+1}(\vsigma)=( p(\rho_{g,1}^{n+1}(\vsigma_1)), p(\rho_{g,2}^{n+1}(\vsigma_2)),\dots, p(\rho_{g,M}^{n+1}(\vsigma_M)),
\vsigma_{M+1},\dots,\vsigma_N)^T.
\end{array}
\end{equation}
Note that $\vecM$ can be written as
\begin{equation}
\vecM=\begin{pmatrix}
{\bf 1} & \vz \\\vecM_1 & \vecM_2
\end{pmatrix}
\end{equation}
where ${\bf 1}$ is the $M\times M$ unity matrix and ${\bf M}_1\in\mathbb{R}^{(N-M)\times M}$, ${\bf M}_2\in\mathbb{R}^{(N-M)\times(N-M)}.$
We abbreviate the third term as $\vecV(\vsigma)$ and obtain similar to \rf{bc_velocity}:
\begin{equation}\label{Velocity}
\begin{array}{ll}
\dt\vecV(\vsigma)=\frac{1}{2}(
&(|\Gamma_1|\v_{g,1}^{n+1}(\vsigma_1)\cdot\vecn_1+|\Gamma_M|\v_{g,M}^{n+1}(\vsigma_{M})\cdot\vecn_M+\tilde{c}_1),\\
&(|\Gamma_2|\v_{g,2}^{n+1}(\vsigma_2)\cdot\vecn_2+|\Gamma_1|\v_{g,1}^{n+1}(\vsigma_{1})\vecn_1+\tilde{c}_2),\\
&\dots,\\
&(|\Gamma_M|\v_{g,M}^{n+1}(\vsigma_M)\vecn_M+|\Gamma_{M-1}|\v_{g,M-1}^{n+1}(\vsigma_{M-1})\vecn_{M-1}+\tilde{c}_M),\\
&0,~\dots~,0).
\end{array}
\end{equation}
We multiply \eqref{operator} with $\dt$ and obtain
 \begin{equation}\label{proofoperator}
 \dt\ds\nabla_\vsigma\cdot \vecO(0)=\dt{\bf M}\nabla_\vsigma\cdot {\bf P}(\vsigma)+\nabla_\vsigma\cdot{\bf V}(\vsigma).
 \end{equation}
 A short computation (for $\dx=\dy$) yields
 \begin{equation}
 \begin{array}{lll} 
 \nabla_\vsigma\cdot{\bf P}(\bf{0})&=&\left(\begin{array}{cc}
 {\bf A}&0\\0&\bf{1}
 \end{array}\right)\\[2ex]
  \nabla_\vsigma\cdot{\bf V}^d(\bf{0})&=&\left(\begin{array}{cc}
 {\bf B}&0\\0&0
 \end{array}\right)
  \end{array}
 \end{equation}
 with
 \begin{equation}
 \begin{array}{l}
 {\bf A}=\ds\frac{1}{\rho_l}\left(\begin{array}{ccccc}
 a_1&0&\cdots&\cdots&0\\
 0&a_2&\ddots&&\vdots\\
 \vdots&\ddots&\ddots&\ddots&\vdots\\
  \vdots&&\ddots&\ddots&0\\
 0&\cdots&\cdots&0&a_M
 \end{array}\right),\quad
  {\bf B}=\frac{\dx}{2}\left(\begin{array}{ccccc}
 b_1&0&\cdots&0&b_M\\
 b_1&b_2&0&\cdots&0\\
 0&\ddots&\ddots&&\vdots\\
  \vdots&&\ddots&\ddots&0\\
 0&\cdots&0&b_{M-1}&b_M
 \end{array}\right)
 \end{array}
 \end{equation}
 and ${\bf 1}$ the $(N-M)\times(N-M)$ unity matrix.
 The values $a_K,b_K,~K=1,\dots,M$ are computed by the derivatives of the lax curves 
 \begin{equation}
 \left(\begin{array}{c}a_K\\b_K 
 \end{array}\right)=\partial_{\vsigma_K}{L}_i^g(0,\rho_{g,K},\v_{g,K})=\left(\begin{array}{c}\rho_{g,K}^0\\(-1)^ia 
 \end{array}\right),\quad i=1,2.
 \end{equation}
 We can easily see that ${\bf A}$ has full rank and so has $\vecM$. After these preparations we can write \rf{proofoperator} as
 \begin{equation}\label{proof1}
 \begin{array}{ll}
 \dt\nabla_\vsigma\cdot \vecO(\bf{0})&=\dt\vecM\left(\begin{array}{cc}
 {\bf A}&0\\0&\bf{1}
 \end{array}\right)+\left(\begin{array}{cc}
 {\bf B}&{\bf 0}\\{\bf 0}&{\bf 0}
 \end{array}\right)\\[2ex]
 &=\left({\begin{array}{cc}
  \dt\vecA+\vecB&{\bf 0}\\\dt\vecM_1\vecA&\dt\vecM_2
 \end{array}}\right).
  \end{array}
 \end{equation}
 To compute the determinant of $\dt\nabla_\vsigma\cdot \vecO(\vsigma)$ we use the following result for the determinant  of a matrix 
 \[\vecD=\left(\begin{array}{cc}
 \vecD_1&\vecD_2\\\vecD_3&\vecD_4
 \end{array}\right)\]  
 formed by 4 block matrices. 
 If $\vecD_4$ is invertible we have
 \begin{equation}
 \text{det}(\vecD)=\text{det}(\vecD_4)\cdot\text{det}(\vecD_1-\vecD_2\vecD_4^{-1}\vecD_3).
 \end{equation}
 In our case we compute for $\dt$ sufficiently small
 \begin{equation}
 \begin{array}{ll}
 \dt\nabla_\vsigma\cdot \vecO(\bf{0})&=\dt\cdot\text{det}(\vecM_2)\cdot\text{det}(\dt\vecA+\vecB)\\
 \end{array}
 \end{equation}
 As the mass matrix $M$ has full rank so does $\vecM_2$ and it remains to proof that $\text{det}(\dt\vecA+\vecB)\neq0$.
A short computation yields
\begin{equation}
\begin{array}{ll}
\text{det}(\dt\vecA+\vecB)&=-\ds\left(\frac{a\dx}{2}\right)^M+\prod_{K\in\mathcal{I}_l^{nodes}}\left(\frac{a\dx}{2}+\frac{\dt}{\rho_l} \rho^0_{g,K}\right)\\[2ex]
&=\ds\left(\frac{a\dx}{2}\right)^{M-1}\frac{\dt}{\rho_l}\sum_{K\in\mathcal{I}_l^{nodes}}\rho^0_{l,K}+O(\dt^2)\\[2ex]
&\neq0
\end{array}
\end{equation}
The last line is true for $\dt$ small enough because $\rho_{l,K}^0>0$ for all $K\in \mathcal{I}_l$.

\textbf{Remark} This result holds also for a droplet with wall contact. For each boundary edge of a cell $B_I$ at the wall, we add a weakly compressible ghost cell $B_O$ outside the domain with values
\[\v^n_O\cdot\vecn=-\v^n_I\cdot\vecn,\quad p_O^n=p_I^n.\]
The pressure inside the weakly compressible domain is governed by the Tait equation. With the help of this ghost cell ansatz, we are able to obtain a similar result as in Theorem \ref{maintheorem}.

\begin{corollary}$\vecO$ satisfies the conditions of the inverse function theorem \cite[Thm C.8]{Evans-book-10} and thus $\vecO$ has a unique solution in ${\bf 0}$.
\end{corollary}

The coupling \eqref{bc_velocity},\eqref{bc_pressure} exactly fulfils the interface conditions \eqref{boundary} across each interface. From now on we refer to this nonlinear compressible/incompressible coupling as (NCIC). Unfortunately, we have to solve a coupled system of non-linear equations. The solution of this system requires a high computational effort at each time step of the algorithm. Another difficulty arises, if the wave curves $\mathcal{L}^1,\mathcal{L}^2$ are not given in a closed algebraic expression, as it is often the case for the real gases. These difficulties can be resolved by two different approaches. 
\par\medskip
{\bf 1. Linearized compressible/incompressible coupling (LCIC)}
 
 We want to present the approximation via linearised wave curves as an example for the solution strategy with approximate Riemann solvers. More examples and detailed explanations can be found in \cite{JaegleSchleper-10}
 
 The key idea is to replace the wave curves $\mathcal{L}^1,\mathcal{L}^2$ by the tangent curves $\bar{\mathcal{L}}^1,\bar{\mathcal{L}}^2$ around the initial states $\vecu^-,\vecu^+$, respectively. If we consider the ideal gas law in \eqref{Rleft},\eqref{Rright}, these linearised wave curves are given by
 \begin{equation}
 \bar{\mathcal{L}}^1=\begin{pmatrix}\rho^-\\\v^-\cdot\vecn\end{pmatrix}
 +\vsigma\begin{pmatrix}\rho^-\\-a\end{pmatrix},\quad\bar{\mathcal{L}}^2=\begin{pmatrix}\rho^+\\\v^+\cdot\vecn\end{pmatrix}
 +\vsigma\begin{pmatrix}\rho^+\\a\end{pmatrix}
 \end{equation}
 We use these curves to compute $\v^{n+1},~p^{n+1}$ in equations \eqref{bc_velocity},\eqref{bc_pressure}.
 Note that now we only have to  solve a \emph{linear system} instead of a non-linear one. We can easily check, that this system is well defined, if we look at equations \eqref{Pressure}-\eqref{Velocity}. In order to compute $\nabla_\vsigma\cdot\vecP({\bf 0}),\nabla_\vsigma\cdot\vecV({\bf 0})$, 
 we need the values $\partial_\sigma\bar{\mathcal{L}}^1(0),~\partial_\sigma\bar{\mathcal{L}}^2(0)$. As $\bar{\mathcal{L}}^1,~\bar{\mathcal{L}}^2$ are the tangent curves in $\vecu^-,~\vecu^+$, we have $\bar{\mathcal{L}}^i(0)=\mathcal{L}^i(0),~(i=1,2)$. Another benefit of this approach comes from the face that we only need to evaluate the pressure and sound speed once per time step. Note that this coupling does not provide the exact solution to the Riemann problem. However, if the jump in the initial data is not too large, we expect the scheme to provide sufficient accuracy.  
\par\medskip
{\bf 2. Explicit in-time compressible/incompressible coupling (ECIC)}

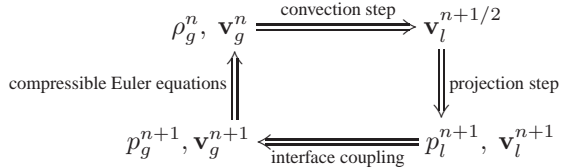
\begin{figure}[ht!]
\begin{xy}
\hspace{2cm}
  \xymatrix{
      **[l]\rho_g^n,~\v_g^n \ar@{=>}[r]^{\text{convection step}} \ar@{<=}[d]_{\text{compressible Euler equations}}    &   **[r]\v_l^{n+1/2} \ar@{=>}[d]^{\text{projection step}}  \\
      **[l]p_{g}^{n+1},\v_{g}^{n+1} \ar@{<=}[r]_{\text{interface coupling}}             &  **[r]p_l^{n+1},~\v_l^{n+1}   
  }
\end{xy}
\caption{Different steps of the explicit in-time compressible/incompressible coupling.}
\label{fig:coupling_circle}
\end{figure}

Instead of computing the boundary values using the wave curves, we prescribe the compressible values $\vecu_g^n$ as boundary values for $\v_l^{n+1},~p_l^{n+1}$. This leads to a smaller linear system as the boundary values do not depend on $\vsigma$. We can regard this system as an elliptic system for $p^{n+1}$ with Dirichlet boundary data. There are various existence and uniqueness results for this system (c.f. \cite{Ciarletbook}). Now we use the boundary values $\v_l^{n+1},p_l^{n+1}$ to compute the boundary values for $\v_g^{n+1},p_g^{n+1}$ at the next time level. Note that this solution strategy violates the coupling condition $\v_l^{n+1}\cdot\vecn=\v_g^{n+1}\cdot\vecn,~p_l^{n+1}=p_g^{n+1}$ and instead fulfils $\v_l^{n+1}\cdot\vecn=\v_g^{n}\cdot\vecn,~p_l^{n+1}=p_g^{n}$.  
Figure \ref{fig:coupling_circle} illustrates the different steps for this coupling.
Note that standard schemes for the incompressible Euler equations only use velocity boundary conditions. However, due to the nonphysical pressure inside the droplet, the backward coupling $p_g^{n+1}=p_l^{n+1}$ did not work for our scheme. With the coupling we outlined above, our numerical experiments showed the expected results.

We refer to Section \ref{s:numerics} for further details and for numerical comparison of the different solution strategies. 

\section{Numerical Experiments}
\label{s:numerics}
In the following we present two numerical tests in two space dimensions to show that the incompressible/compressible system is a good approximation of the compressible/compressible system. The numerical solutions are computed with the numerical schemes (NCIC), (LCIC), (ECIC) that we presented in Section \ref{s:Algorithm}. We compare the results for these numerical schemes with a scheme for the fully compressible problem (CCC).
 \subsection{Planar Interface}
 We consider a rectangular "droplet" that is surrounded by gas in the domain $\Omega=[0,6]\times[0,1]$. 
 This setting corresponds to the initial conditions 
 \begin{equation}
 \begin{array}{l}
 \begin{pmatrix}
 \rho_g(0,\vecx)\\\v_g(0,\vecx)\end{pmatrix}
 =\left\{\begin{array}{ll}
  \begin{pmatrix}
  1.5\\0\\0
  \end{pmatrix},&x\in[0,2),~y\in[0,1],\\
  \begin{pmatrix}
  1.0\\0\\0
  \end{pmatrix},&x\in(3,6),~y\in[0,1],
 \end{array}\right.\\[8ex]
  \begin{pmatrix}
 \rho_l(0,\vecx)\\\v_l(0,\vecx)\end{pmatrix}=
  \begin{pmatrix}
  500\\0\\0
  \end{pmatrix},\quad \quad x\in[2,3],~y\in[0,1].
  \end{array}
  \end{equation}
The initial conditions are constructed such that the setting corresponds to a situation where a shock wave, carrying a pressure jump of $0.5$ hits the droplet at $t=0$. Furthermore we discretize $\Omega$ with $\dx=\dy=0.125$ and choose $a=1$ in the ideal gas law \eqref{ideal} and $k=5000/7,~\rho_0=500,~\gamma=7,~p_0=1$ in the Tait equation \eqref{tait}, respectively. 
Note that this test case is a 2D expansion of a 1D example, that we took from \cite{Schleper}. An analytical convergence result covering this test case can be found in \cite{Colombo-09}.
We keep the right domain boundary as outflow boundary whereas the other 3 boundaries are assumed to be solid walls. This setting yields several rarefaction and shock waves hitting the droplet during the computation time up to $T=600$. 
  \begin{figure}[h!]
  \includegraphics[width=0.8\textwidth, height=7cm]{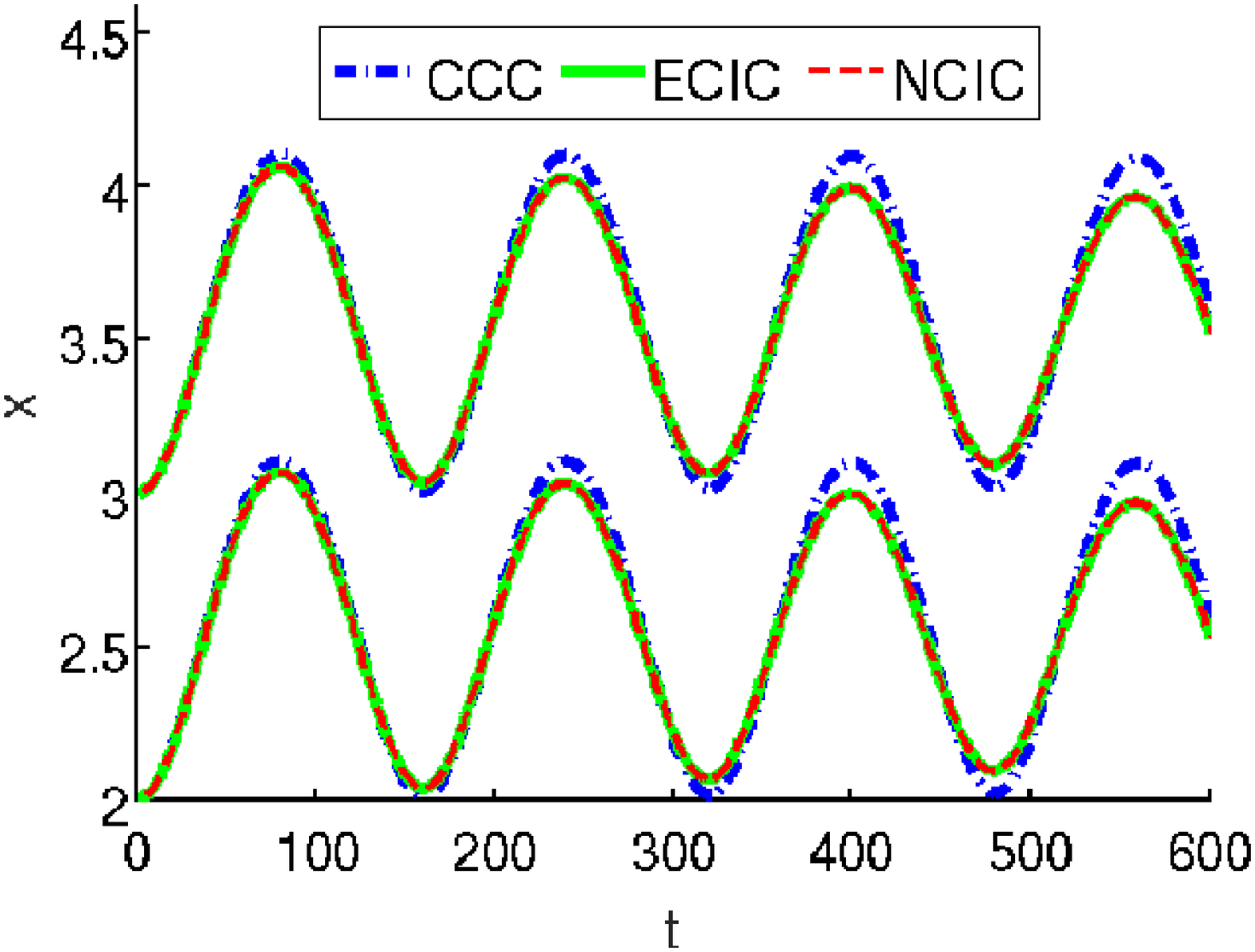}
  \caption{Comparison of the interface location for three different couplings: CCC (dashed dotted line), NCIC (solid line) and ECIC  (dashed line). NCIC and ECIC are visually indistinguishable.}
  \label{fig:planar_interface}
  \end{figure}
Figure \ref{fig:planar_interface} shows the evolution of the interface location for three different numerical schemes. We compare the fully compressible model \eqref{gEuler}-\eqref{ccboundary} and the mixed model \eqref{cEuler}-\eqref{boundary} with the implicit discretization as well as with the explicit discretization. We can see that the solutions are close, but the amplitude for the compressible/incompressible models decreases with increasing computation time, whereas the oszillation frequencies for the three schemes are in very good agreement. This behaviour can be explained by the fact, that we have a weakly compressible liquid for the CCC and an incompressible one for both the NCIC and ECIC.
  
In Table \ref{tab:CPU_planar} we report the CPU time for the solution of this test case. One can see the number of time steps is reduced by the use of the coupled model. However the implicit interface discretization is computationally very expensive so that the advantage is only small. In contrast, the CPU time for the explicit coupled model is reduced drastically. Note that the performance of the coupled models compared to the fully compressible model depends on the ratio of the compressible and weakly compressible speeds of sound. If this ratio is large, the time saving is large as well. 
\begin{table}[h!]
\centering
\begin{tabular}{|c||c|c|c|}
\hline
&$t_c$&$t_c/t_{c,CCC}$&time-steps\\\hline\hline
CCC&2323&1.0&28232\\\hline
NCIC&1751&0.7538&8999\\\hline
ECIC&672&0.2893&8999\\\hline
\end{tabular}
\caption{Comparison of the different models with respect to the computational effort: CPU time ($t_c$) and number of time steps.}
\label{tab:CPU_planar}
\end{table}
 \subsection{Droplet}
 In this test case we consider the interaction of a spherical droplet with a compressible shock. The purpose of this test is to compare the performance of the coupled equations for different coupling mechanisms. We make some simplifications, because we want to focus on the coupling between the two fluids. 
Firstly, we assume that we have a weak shock which is neither able to deform nor to accelerate the droplet significantly. This means that we have a fixed phase boundary and do not need to track the position of the interface in each time step. This is reasonable, because the densities of gas and liquid differ by a factor of $1000$. 
Secondly, we neglect surface tension effects. As we have a spherical droplet with fixed radius $R$, the curvature is just a constant number in the coupling condition \eqref{boundary}$_2$ and we set $\kappa=0$ for simplicity. 
Thirdly, we recall that the interface is aligned with the cell boundaries. Therefore we can associate each cell with the liquid or the gas phase, respectively. 

The equation of state for the gas phase is given by the ideal gas law \eqref{ideal} with parameter $\gamma=1.4$. The behaviour of the liquid inside the droplet is modelled by the Tait equation \eqref{tait} with parameters $\gamma=7.15,~k_0=3310,~\rho_0=1000,~p_0=1.0$.
The initial conditions for a droplet radius $R=0.00175$ and a shock position $\vecx_s=(-0.002,0)$ are:
 \begin{equation}
 \begin{array}{l}
   \begin{pmatrix}
 \rho_l(0,\vecx)\\\v_l(0,\vecx)\end{pmatrix}=
  \begin{pmatrix}
  1000\\0\\0
  \end{pmatrix},\quad \quad \vecx\in B_R({\bf 0}),\\[4ex]
  \begin{pmatrix}
 \rho_g(0,\vecx)\\\v_g(0,\vecx)\end{pmatrix}
 =\left\{\begin{array}{ll}
  \begin{pmatrix}
  1\\0\\0
  \end{pmatrix},&\vecx\in[-0.006,0.002]\times[-0.006,0.006]\setminus B_R({\bf 0}),\\
  \begin{pmatrix}
  1.0\\-\log{1.5}\\0
  \end{pmatrix},&\vecx\in(0.002,0.006]\times[-0.006,0.006]\setminus B_R({\bf 0}).
 \end{array}\right.
  \end{array}
  \end{equation}
  The domain is discretized with equidistant cells with grid width $\dx=\dy=0.00005$ and we compute the solutions up to a computation time $T=0.0025 s$. We keep the right domain boundary as inflow boundary and left as outflow boundary, whereas the other 2 boundaries are assumed to be solid walls.
  
\begin{figure}
 \label{fig:pressure}
 \includegraphics[width=0.9\textwidth]{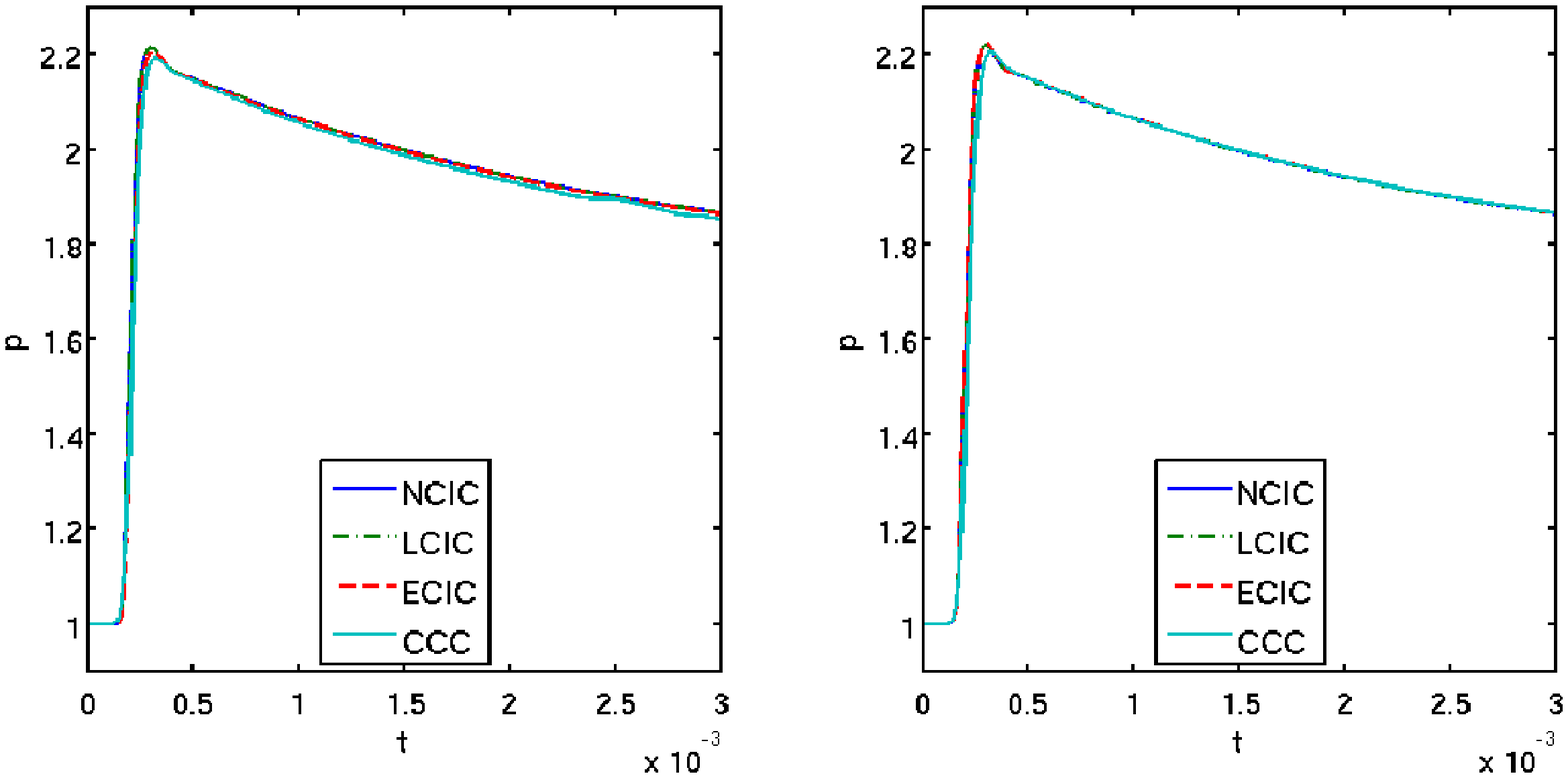}
 \caption{Pressure evolution over time for the different schemes for two neighbouring cells in the liquid domain (left) and the gaseous domain (right) at the point of the interface where the shock wave first hits the droplet.}
 \end{figure}

We compare the results for the fully compressible model and for the three different coupled models that we introduced in Section \ref{s:coupling}. 
\begin{figure}
 \subfigure[NCIC: t=0.0005 s]{\includegraphics[width=0.3\textwidth]{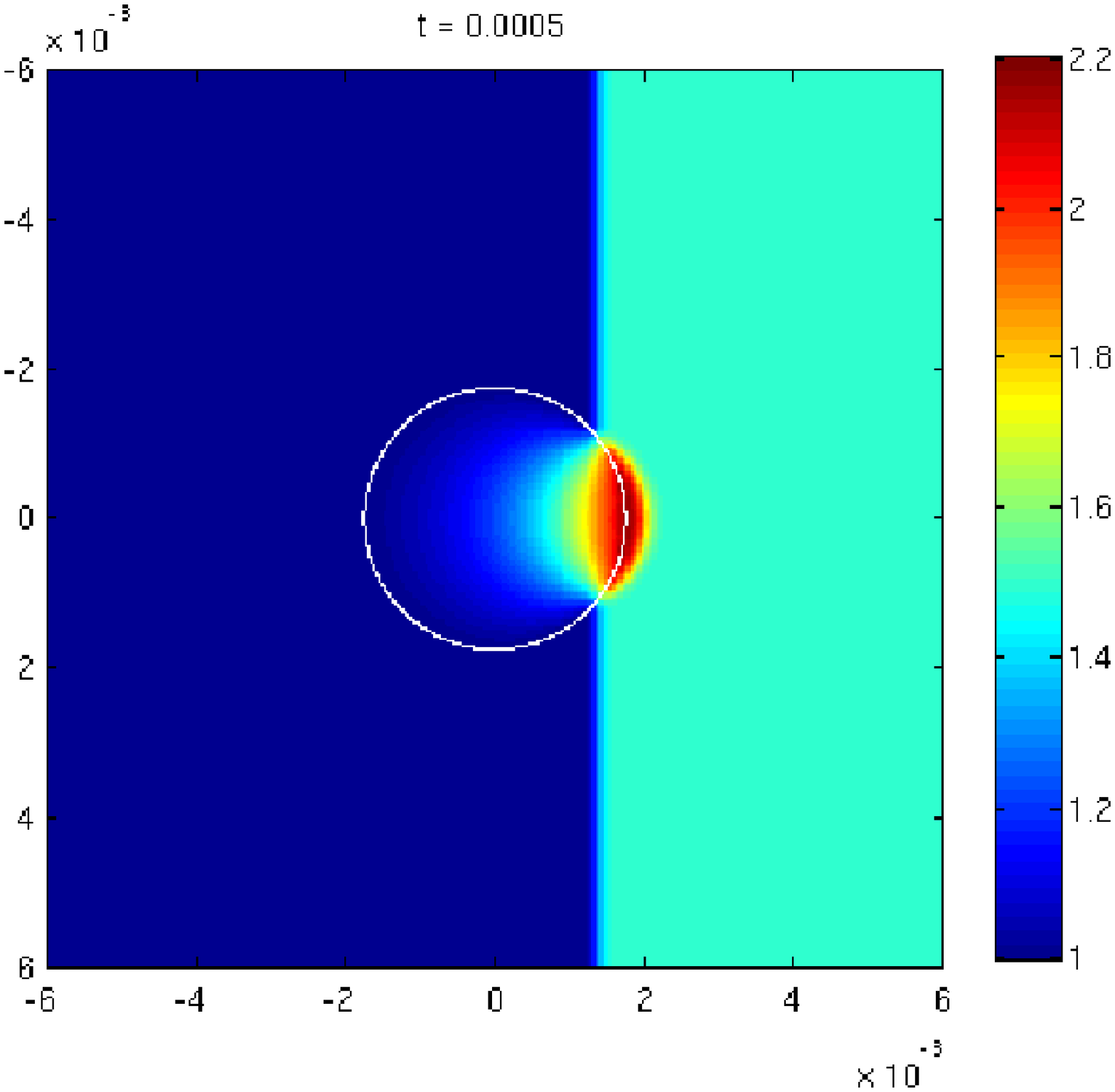}}
 \subfigure[NCIC: t=0.001 s]{\includegraphics[width=0.3\textwidth]{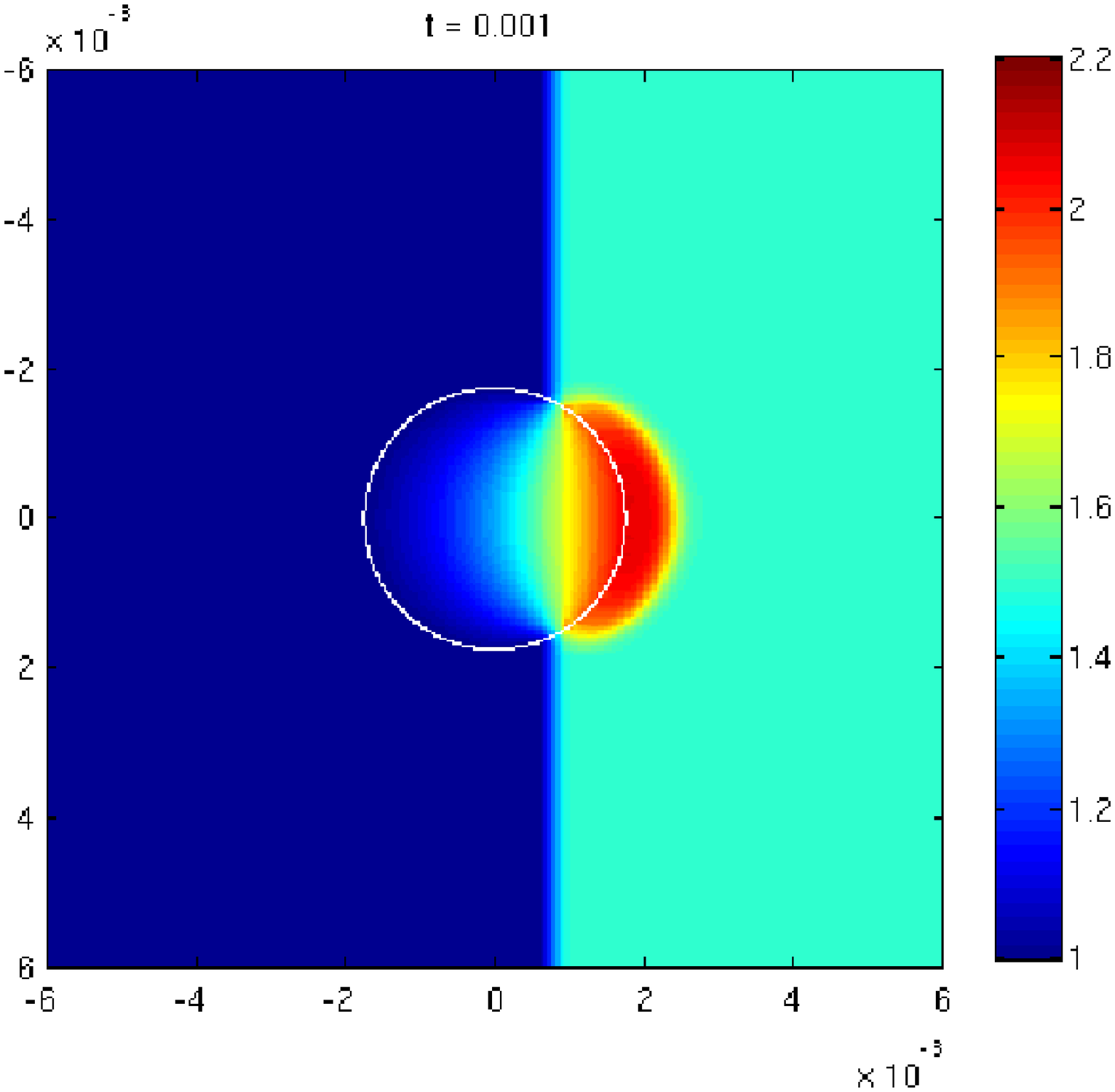}}
 \subfigure[NCIC: t=0.0015 s]{\includegraphics[width=0.3\textwidth]{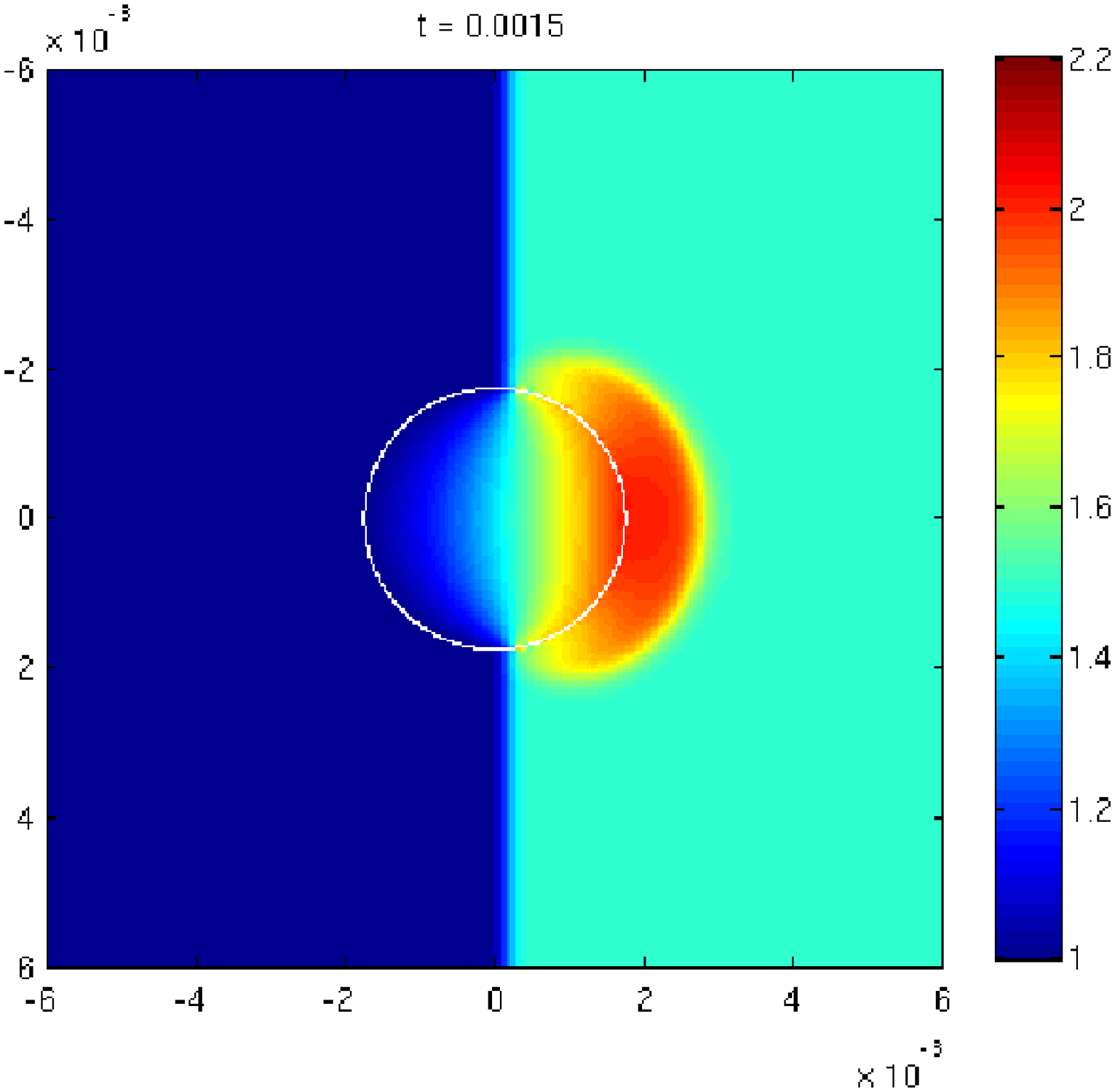}}\\
  \subfigure[LCIC: t=0.0005 s]{\includegraphics[width=0.3\textwidth]{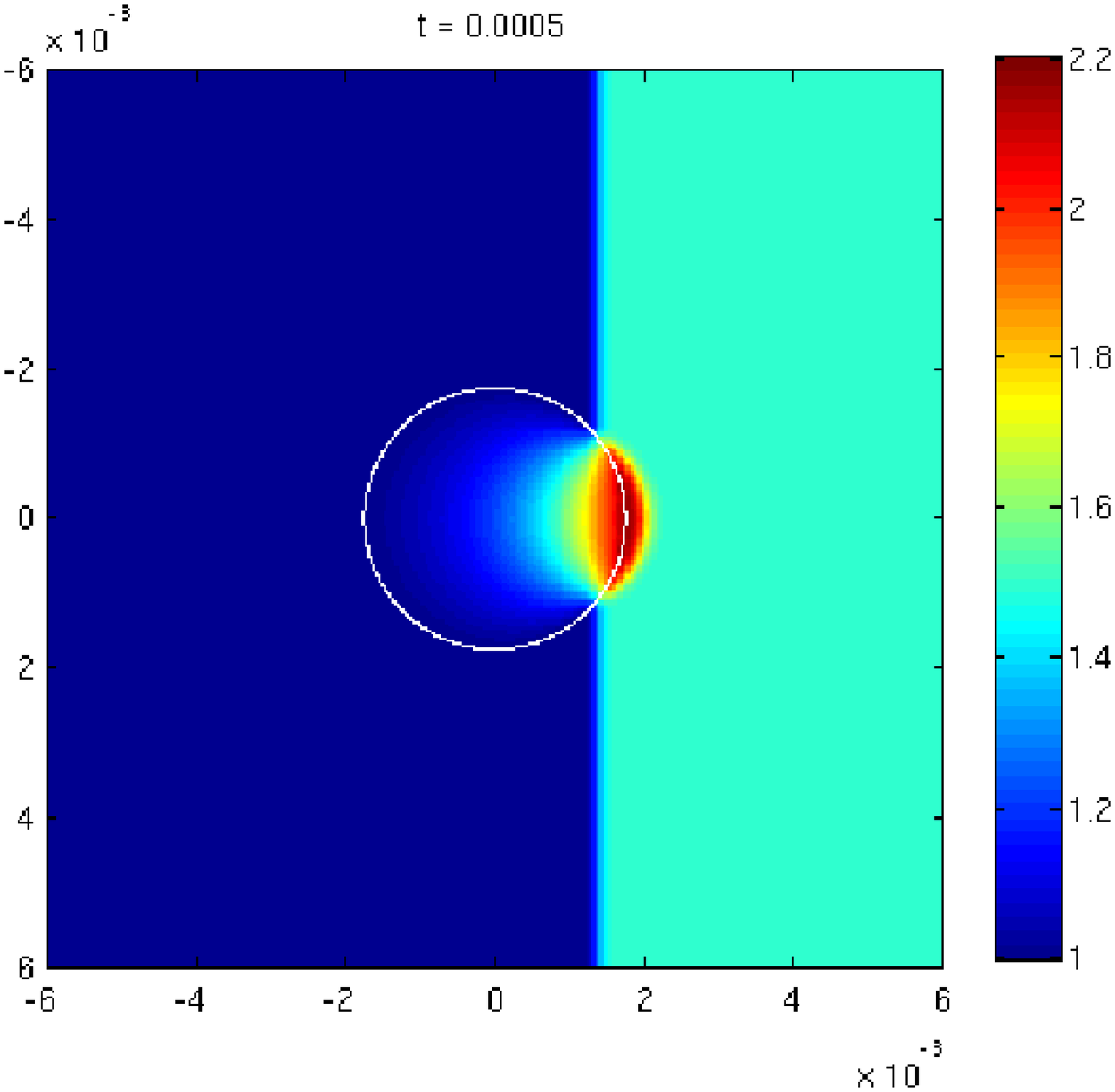}}
 \subfigure[LCIC: t=0.001 s]{\includegraphics[width=0.3\textwidth]{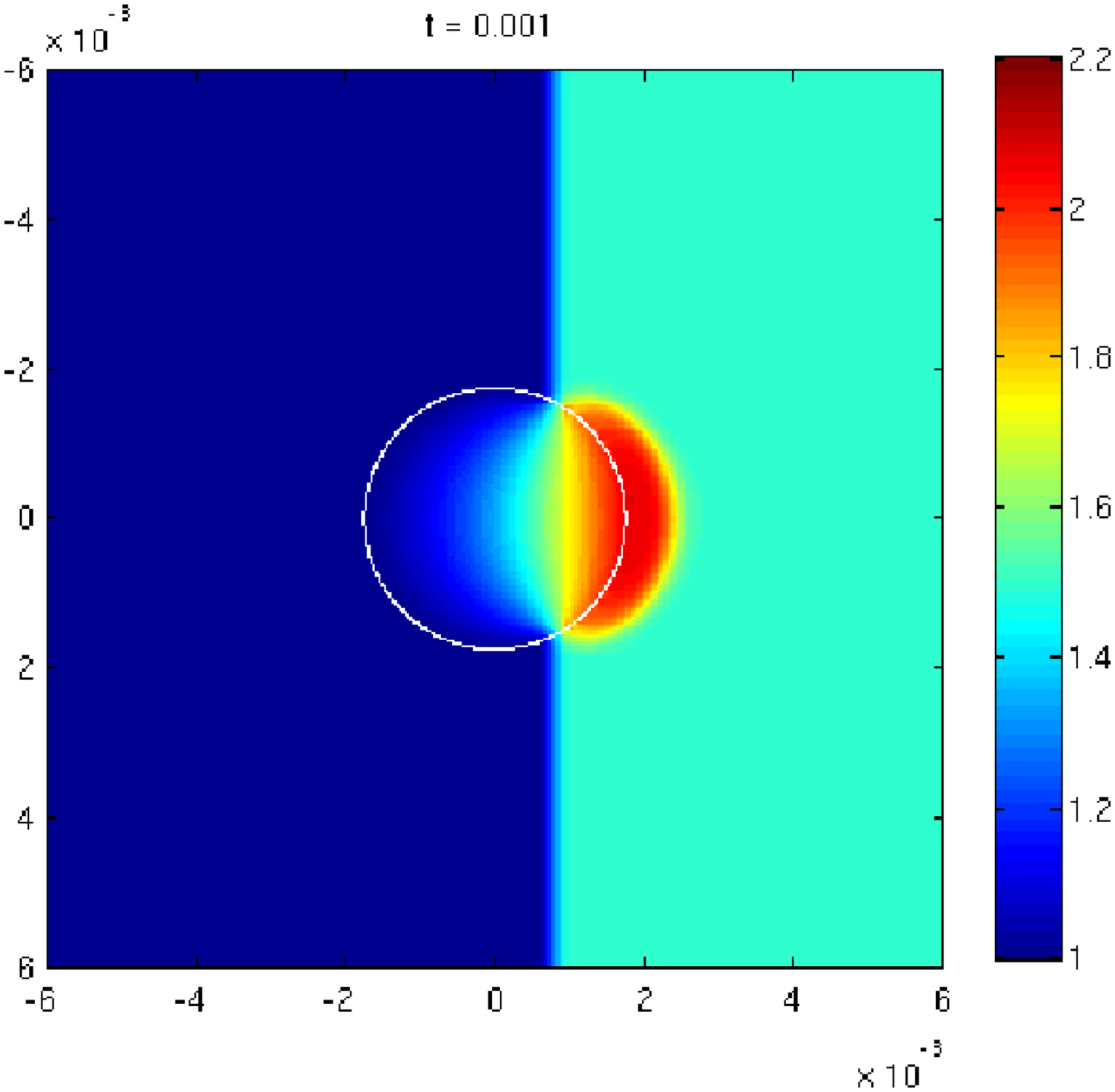}}
 \subfigure[LCIC: t=0.0015 s]{\includegraphics[width=0.3\textwidth]{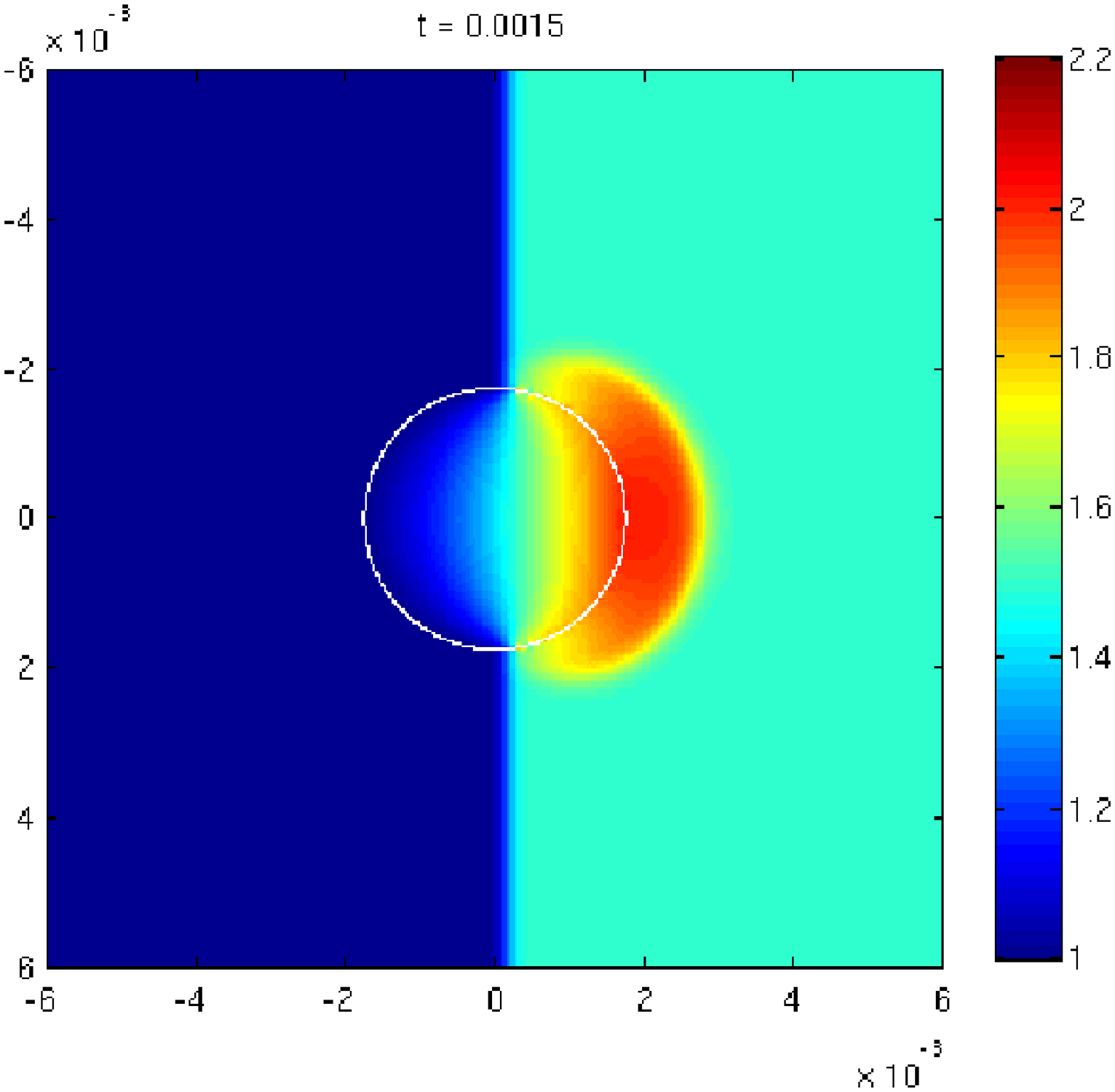}}\\
 \subfigure[ECIC: t=0.0005 s]{\includegraphics[width=0.3\textwidth]{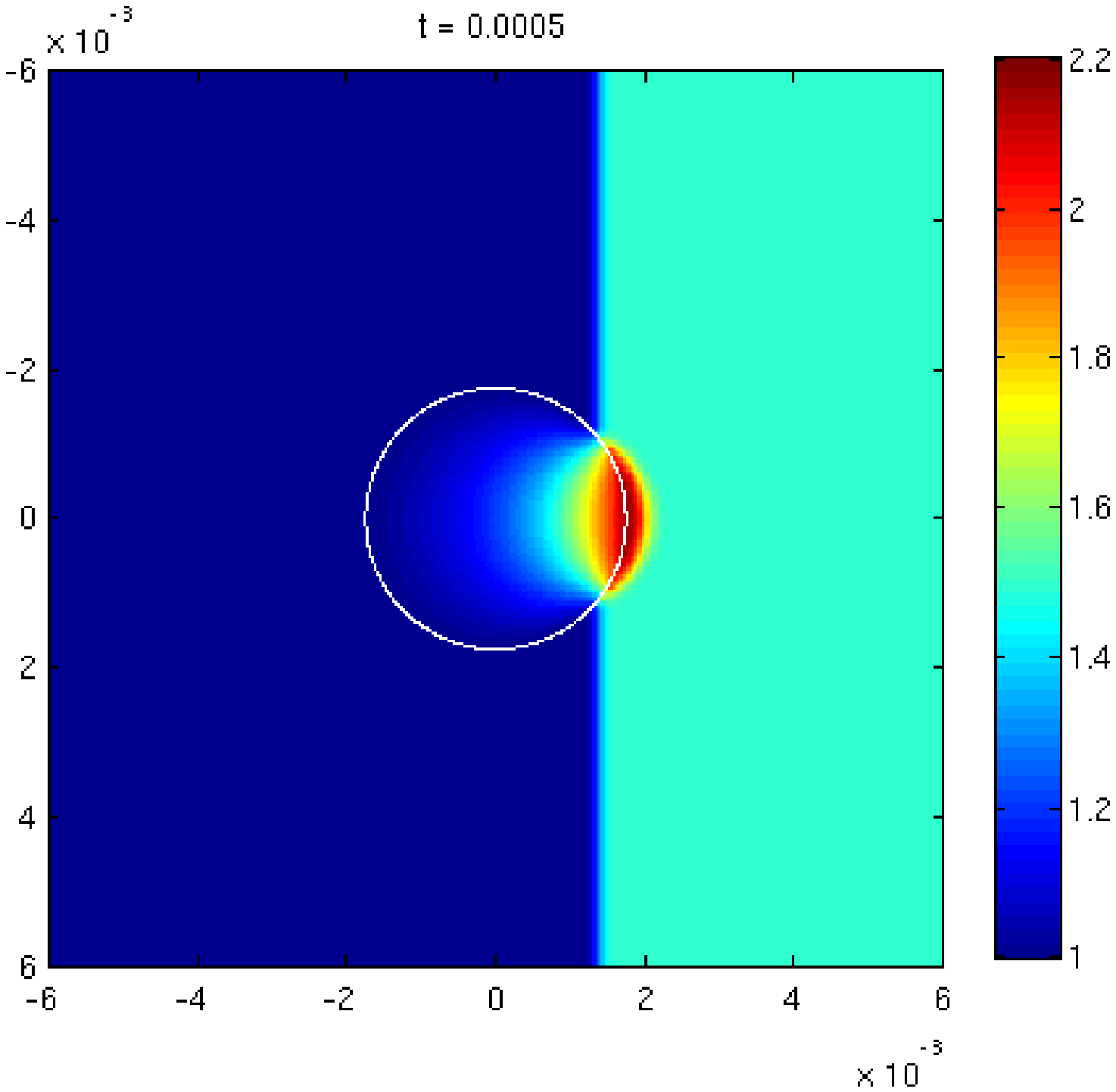}}
 \subfigure[ECIC: t=0.001 s]{\includegraphics[width=0.3\textwidth]{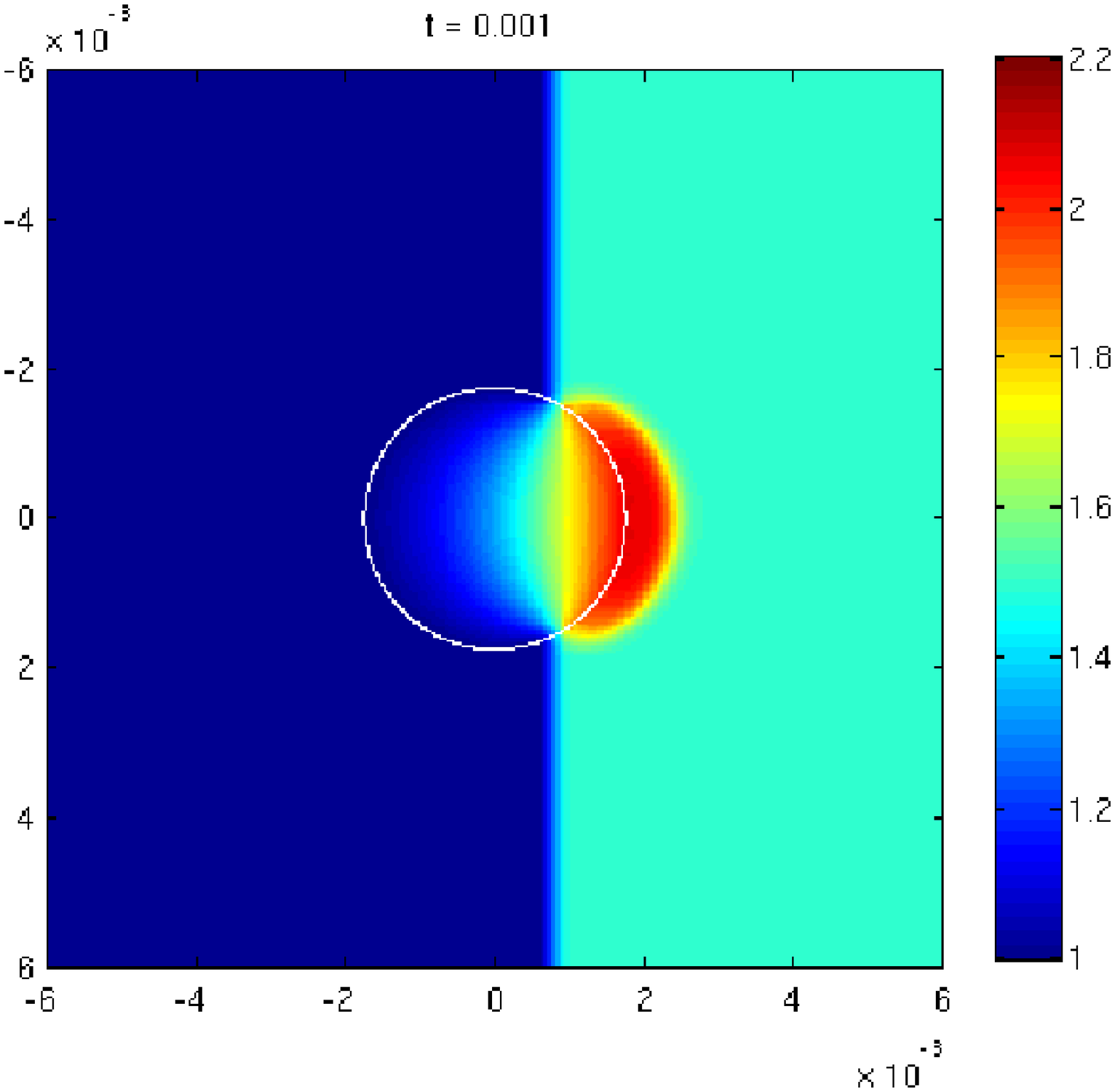}}
 \subfigure[ECIC: t=0.0015 s]{\includegraphics[width=0.3\textwidth]{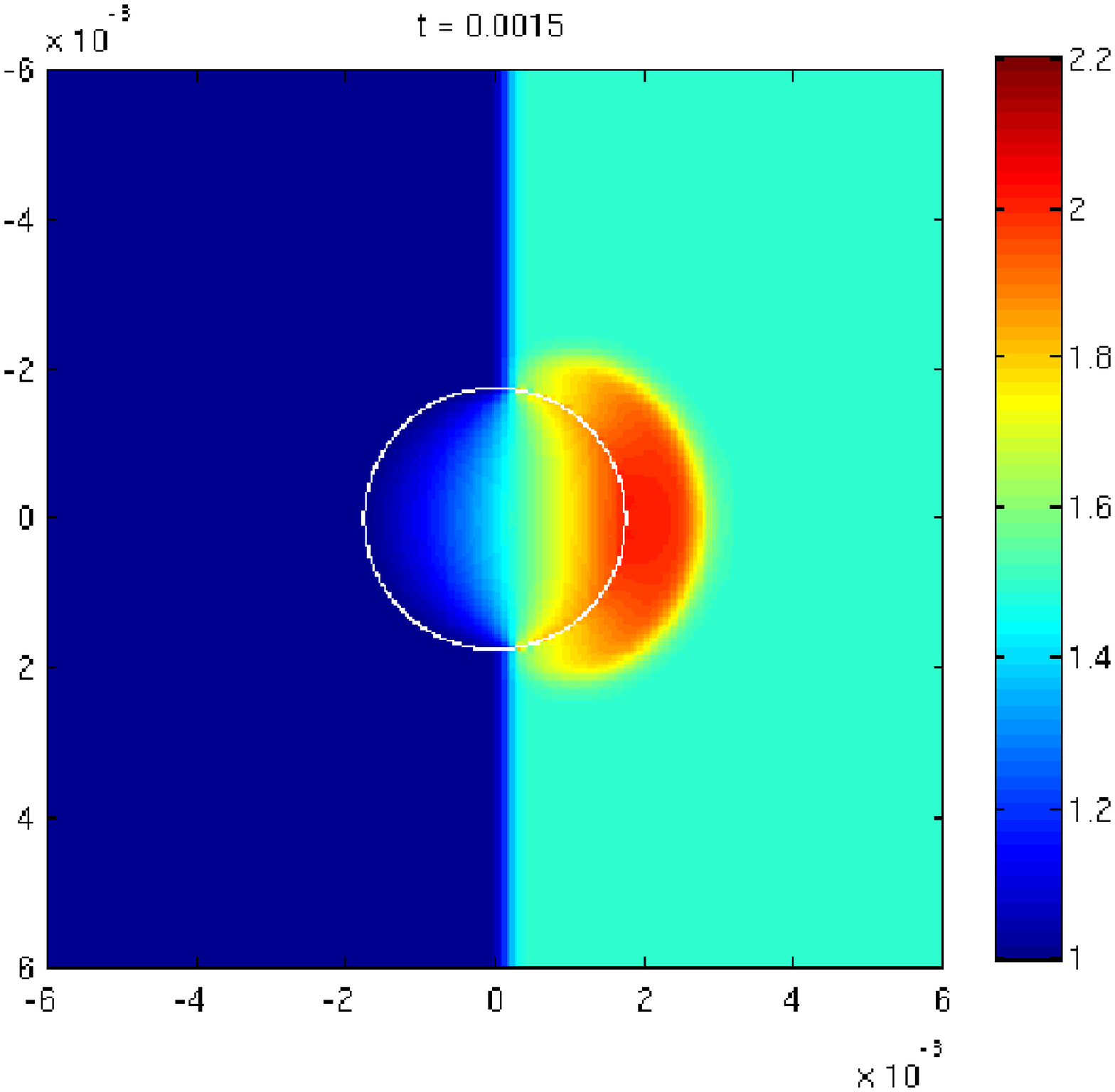}}\\
 \subfigure[CCC: t=0.0005 s]{\includegraphics[width=0.3\textwidth]{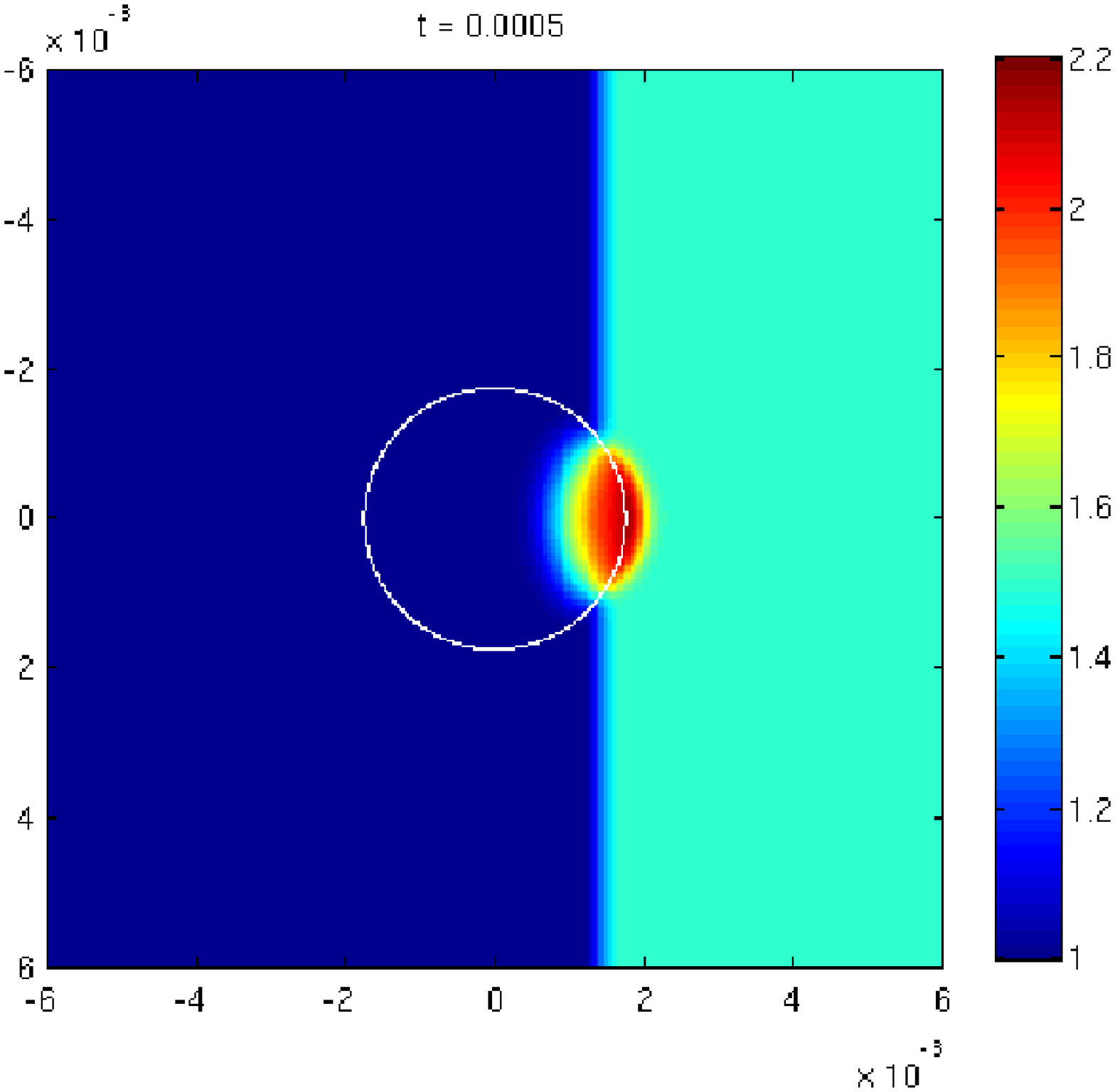}}
 \subfigure[CCC: t=0.001 s]{\includegraphics[width=0.3\textwidth]{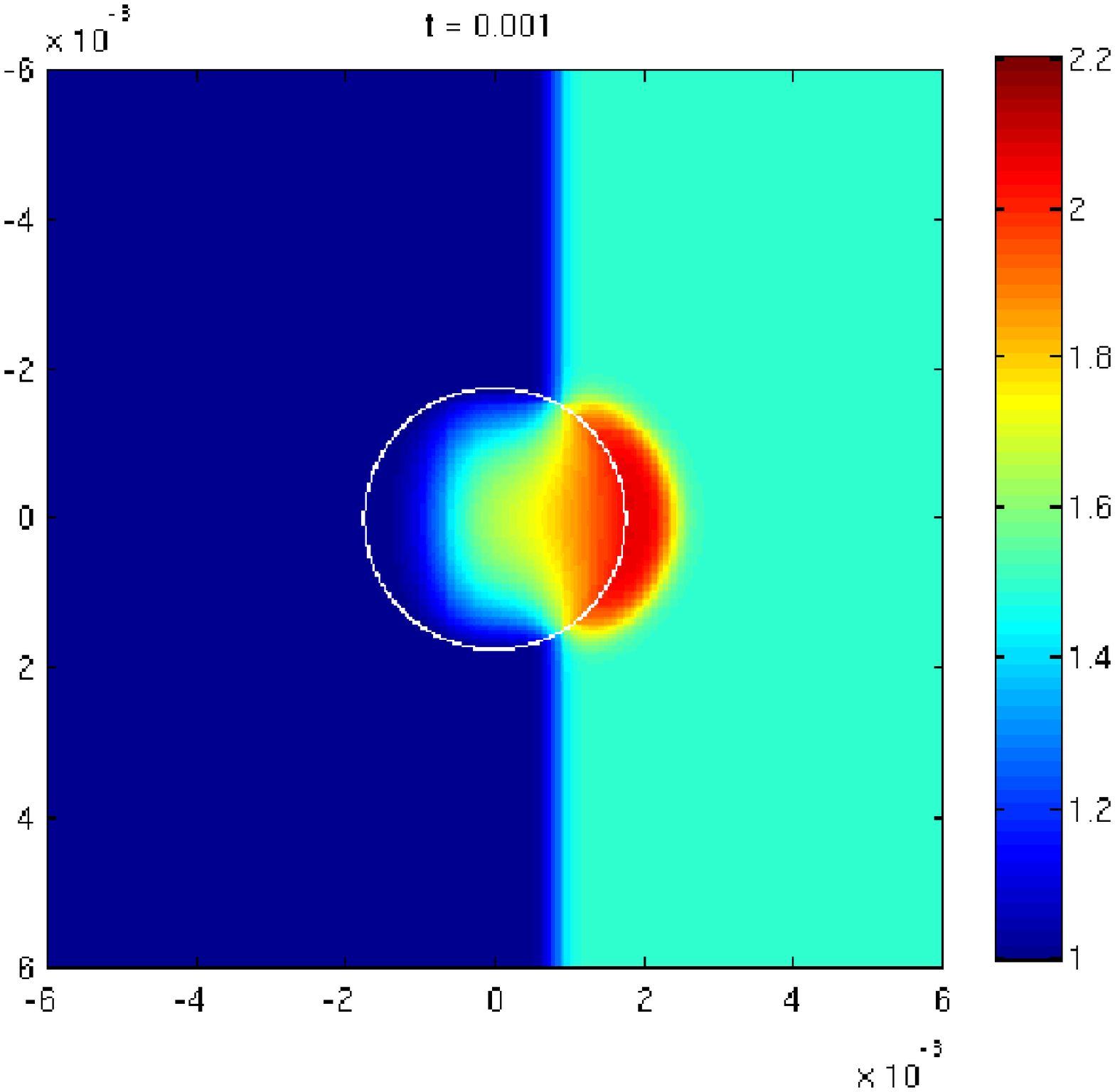}}
 \subfigure[CCC: t=0.0015 s]{\includegraphics[width=0.3\textwidth]{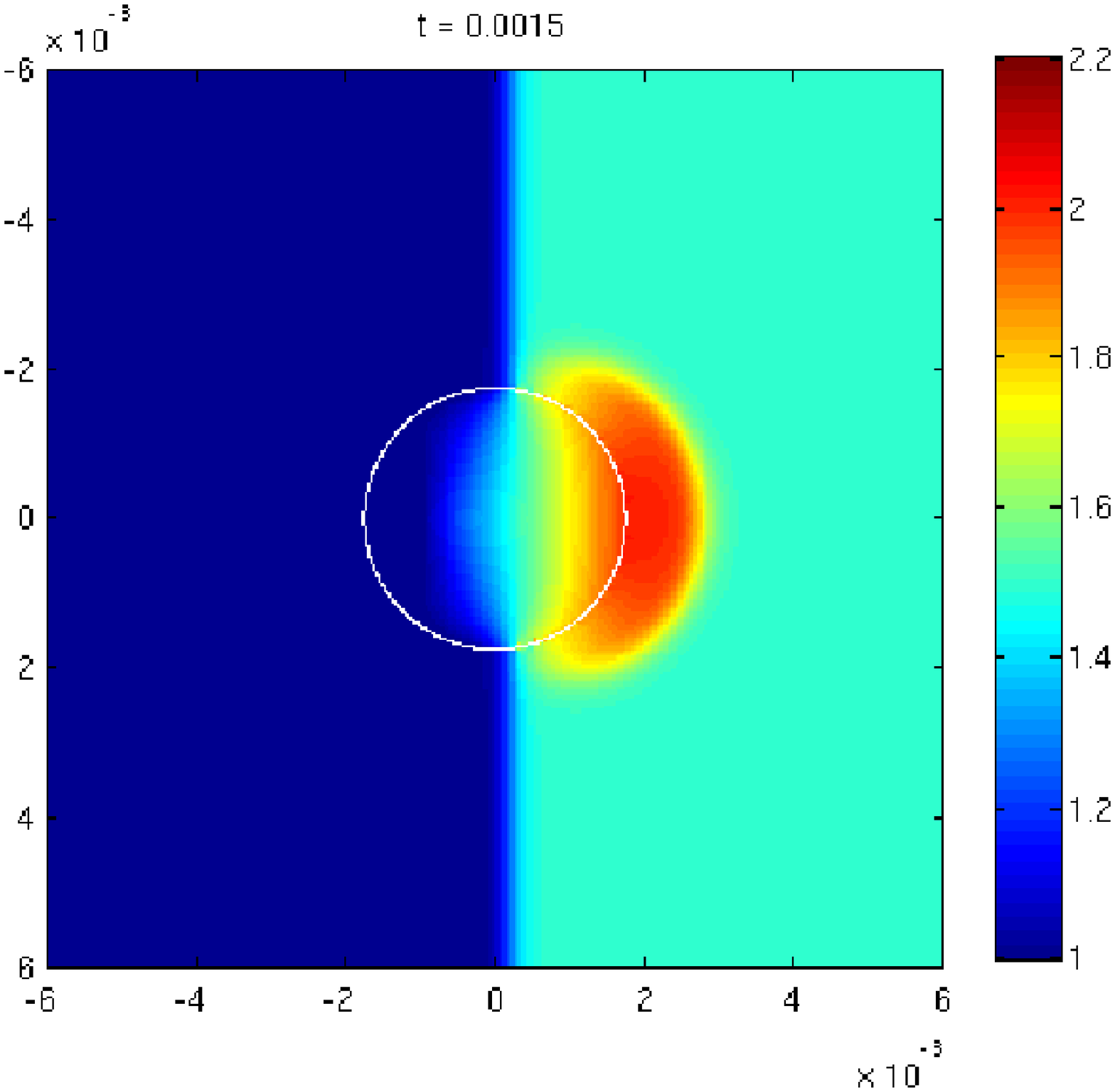}}\\
\caption{Pressure distribution for the three different couplings of the compressible-incompressible Euler equations}
\label{fig:droplet}  
 \end{figure}  
Figure \ref{fig:droplet} shows the results at different time levels with respect to the pressure $p$. The four different solutions reproduce the expected physical behaviour. The incoming shock is partly reflected of the droplet surface whereas the other part enters the liquid. The shock wave propagates faster inside the droplet because of the larger speed of sound. Inside the droplet, the pressure of the fully compressible system differs from the pressures for the coupled systems, whereas the pressures in the gaseous domain are similar for all for systems. This statement is undermined by Figure \ref{fig:pressure}, where the pressure for two neighbouring boundary cells both in the liquid and the gaseous domain at the point of the interface where the shock wave first hits the droplet is plotted over time. 
We observe that the pressures for the coupled systems inside the droplet are smaller than the pressure for the compressible system for all times. The solutions for the implicit coupled system and the linear implicit coupled system are very close, whereas the explicit solution underestimates the pressure due to the violation of the interface conditions.
In the gaseous domain, the pressures of the coupled systems are larger at the beginning of the simulation, but converge to the pressure of the compressible system as $t$ increases. 
This means that the shock is stronger reflected by the droplet for the coupled system than it is for the fully compressible system. This behaviour is in agreement with the properties of the different models. In the fully compressible model, the droplet is weakly compressible a larger part of the shock can enter the liquid domain. In the coupled model the droplet is incompressible and a larger part of the shock is reflected.

We conclude this section with the comparison of the CPU times for the different coupled systems and and for different grid widths. 
\begin{table}[h!]
\centering
\begin{tabular}{|c||c|c||c|c||c|c|}
\hline
&\multicolumn{2}{|c||}{0.001}&\multicolumn{2}{|c||}{0.0005}&\multicolumn{2}{|c|}{0.00025}\\\cline{2-7}
&$t_c$&$t_c/t_{c,CCC}$&$t_c$&$t_c/t_{c,CCC}$&$t_c$&$t_c/t_{c,CCC}$\\\hline\hline
CCC&181.2&1.0&993&1.0&4850&1.0\\\hline
NCIC&13.6&0.0751&138&0.1390&4224&0.8709\\\hline
LCIC&9.8&0.0541&115&0.1158&2416&0.4981\\\hline
ECIC&12.9&0.0712&125&0.1259&2370&0.4887\\\hline
\end{tabular}
\caption{Comparison of the different models with respect to the CPU time ($t_c$).}
\label{tab:CPU_droplet}
\end{table}  
Table \ref{tab:CPU_droplet} allows several conclusions for this test case. 
First, the numerical schemes for the incompressible-compressible system are computationally more efficient that the  scheme for the fully compressible system. This was the main reason for the proposition of the new scheme and we see that the scheme shows the desired behaviour. However, the difference  in  the CPU-time for the compressible and the implicit scheme decreases as the gridwidth does. This can be easily explained if we take a closer look at the properties of the two schemes.
The time steps for both schemes scale with $\dt\approx\mathcal{O}(\dx)$. The number of interface edges scales with $\mathcal{O}(\dx^{-1})$ and the number of cells in the incompressible droplet scales with $\mathcal{O}(\dx^{-2})$.
This means that for the fully compressible system we have to solve $\mathcal{O}(\dx^{-1})$ non linear equations at each time step . 
For the implicit scheme we have to solve a non linear system with $\mathcal{O}(\dx^{-2})$ equations. The solution of this non linear system makes the implicit scheme inefficient for small grid widths. 
Thus we can identify different ranges of application for the two different schemes. If we want to have finely resolved solution and have enough CPU-time available, we choose the fully compressible scheme. 
If we are interested in fast results or the number of cells in the incompressible phase is small compared to the number of interface edges, we choose the implicit scheme.

Secondly, Table \ref{tab:CPU_droplet} shows that the linear implicit and the explicit scheme are faster that the implicit scheme. For both schemes we only have to solve a linear system at each time step. For larger grid widths, the difference is not so large and we choose the exact coupling that is provided by the implicit scheme. For small grid widths, the two schemes provide the solution much faster. As the two approximate solvers need approximately the same CPU-time we recommend to use the linear implicit solver, because the compressible and the incompressible cells are coupled at the same time level.

\section*{Acknowledgement}
The authors acknowledge the support by the German Research Foundation
(DFG) in the framework of the Collaborative Research Center Transregio 75 Droplet Dynamics
under Extreme Ambient Conditions and by the Elite program for postDocs of the Baden-W\"urttemberg
Foundation.

\end{document}